\documentclass[preprint,superscriptaddress]{revtex4-1}
\usepackage[utf8]{inputenc}
\usepackage{amsfonts}
\usepackage{graphicx}
\graphicspath{{Pictures/}}
\usepackage{bm} 
\usepackage{amssymb} 
\usepackage{amsmath} 
\usepackage{braket} 
\usepackage{natbib} 
\usepackage{hyperref}
\usepackage[T1]{fontenc}
\usepackage{xcolor}
\usepackage{marginnote}

\newcommand{\vk}{\ensuremath{\mathbf{k}}}
\newcommand{\vq}{\ensuremath{\mathbf{q}}}

\begin{document}

\title{Emergent Charge Order from Correlated Electron-Phonon Physics in Cuprates}

\author{S. Banerjee}
\affiliation{Theoretical Physics III, Center for Electronic Correlations and Magnetism, Institute of Physics, University of Augsburg, 86135 Augsburg Germany,}
\author{W. A. Atkinson}
\affiliation{Department of Physics and Astronomy, Trent University, Peterborough, Ontario K9L 0G2, Canada}
\author{A. P. Kampf}
\affiliation{Theoretical Physics III, Center for Electronic Correlations and Magnetism, Institute of Physics, University of Augsburg, 86135 Augsburg Germany,}

\date{\today}
\begin{abstract}
Charge-density wave order is now understood to be a widespread feature of underdoped cuprate high-temperature superconductors, although its origins remain unclear. While experiments suggest that the charge-ordering wavevector is determined by Fermi-surface nesting, the relevant sections of the Fermi surface are featureless and provide no clue as to the underlying mechanism. Here, focusing on underdoped YBa$_2$Cu$_3$O$_{6+x}$, we propose a scenario that traces the charge-density wave formation to the incipient softening of a bond-buckling phonon. The momentum dependence of its coupling to the electrons in the copper-oxygen planes favourably selects the incommensurate and axial ordering wavevector found in experiments.  But, it requires strong electronic correlations via their cuprate-specific renormalization of the weight and the dispersion of quasiparticles to enable a unique enhancement of the charge susceptibility near the B$_{1g}$-phonon selected wavevector. The frequency of the B$_{1g}$ phonon softens by a few percent, and a lattice instability with concomitant finite-range charge-density wave correlations will form locally, if nucleated by defects or dopant disorder. These results offer the perspective that the complex phase diagram of underdoped cuprates cannot be understood in the context of strong electronic correlations alone. 
\end{abstract}

\maketitle

\section{Introduction}

 The discovery of charge-density wave (CDW) order~\cite{Comin123} in underdoped cuprates raised the question of whether it is intimately related to pseudogap physics~\cite{PhysRevB.95.224511,PhysRevB.95.054518,PhysRevB.94.205117,Efetov2013,PhysRevB.89.184515,Atkinson2015}, and thereby yet another signature of strong electronic correlations. Charge modulations with a moderate correlation length are detected by nuclear magnetic resonance~\cite{Wu2011,Wu2013,Wu2015}, scanning tunneling microscopy~\cite{Comin390,NetoS393,Kohsaka1380}, and x-ray techniques~\cite{Ghiringhelli821,Comin390,NetoS393,Chang2012, PhysRevLett.110.137004}, with the latter finding incommensurate wave-vectors near $\vq_{CO} = 0.3$ reciprocal lattice units, oriented along the crystalline axes with either uniaxial or biaxial character. The CDW wavevector $\vq_{CO}$ continuously drops with increasing hole doping in YBa$_2$Cu$_3$O$_{6+x}$ (YBCO) and Bi$_2$Sr$_{2-x}$La$_x$CuO$_{6+x}$ (Bi-2201)~\cite{Comin123,Comin390,PhysRevLett.110.137004,PhysRevB.94.184511}, whereas a completely opposite trend is observed in charge-stripe ordered La-based 214 cuprates such as La$_{2-x}$Sr$_x$CuO$_4$ (LSCO) and La$_{2-x}$Ba$_x$CuO$_4$ (LBCO)~\cite{PhysRevB.83.104506}. Not only the concomitant spin-stripe order marks 214 cuprates to be distinctly different from YBCO or Bi-2201; also the tilt-pattern of oxygen octahedra, specifically in the low-tempertaure tetragonal (LTT) phase, points to subtle structural differences, since the LTT tilt pattern breaks the four-fold symmetry in the copper-oxygen planes and thereby offers a structure related source for a uniaxial character of the density waves. Additionally, the orbital symmetry of CDW order in YBCO and La$_{1.875}$Ba$_{0.125}$CuO$_4$ was found to be dissimilar~\cite{Achkar2016}. Due to these apparent differences we focus subsequently on the charge ordering in YBCO. 
 
 Stripe formation and the competition of stripe states with d-wave superconductivity or pair-density waves was investigated for multiband Hubbard or $t-J$ models with accurate computational tools~\cite{PhysRevLett.113.046402,Zheng1155,Jiang1424}. Yet, finding the charge-stripe ordered ground states proved elusive due to the near degeneracy of the competing states. So far the axial orientation of $\vq_{CO}$ and its doping variation in YBCO and Bi-2201 proved difficult to reconcile with purely electronic model calculations~\cite{Huang1161,Zheng1155}. Weak-coupling theories typically predict that $\vq_{CO}$ lies on the Brillouin zone (BZ) diagonal~\cite{Efetov2013,PhysRevB.89.075129,PhysRevB.88.155132} unless the CDW instability is preceded by a Fermi surface reconstruction to form hole pockets~\cite{Atkinson2015}. Based on these evidences, we conjecture that some physics ingredient may still be missing in these model calculations.

 In the continuing search for the origin of charge order in cuprates, we employ further experimental facts. One is the discovery of charge order in overdoped Bi-2201 with an unreconstructed Fermi surface~\cite{Peng2018}. This result poses a question as to whether the CDW is actually tied to the pseudogap phenomena. A second experimental hint relates to the softening or broadening of phonon modes at the CDW wavevector~\cite{PhysRevLett.107.177004,Kim1040} and to giant phonon anomalies near the CDW instability, which all point to a strong electron-phonon (\textit{el-ph}) coupling~\cite{LeTacon2013}. While the phonon dispersion will necessarly react to a charge-density modulation, phonons may also play a key role in CDW formation, for example selecting the ordering wavevector in CDW-susceptible materials~\cite{Eiter2013,PhysRevB.77.165135}. In cuprates, the frequencies of the phonons decrease only weakly, and a continuous softening to zero frequency is likely precluded by quenched disorder~\cite{Campi2015,Wu2015}.

\begin{figure}[t]
\centering
\includegraphics[width=0.90\linewidth]{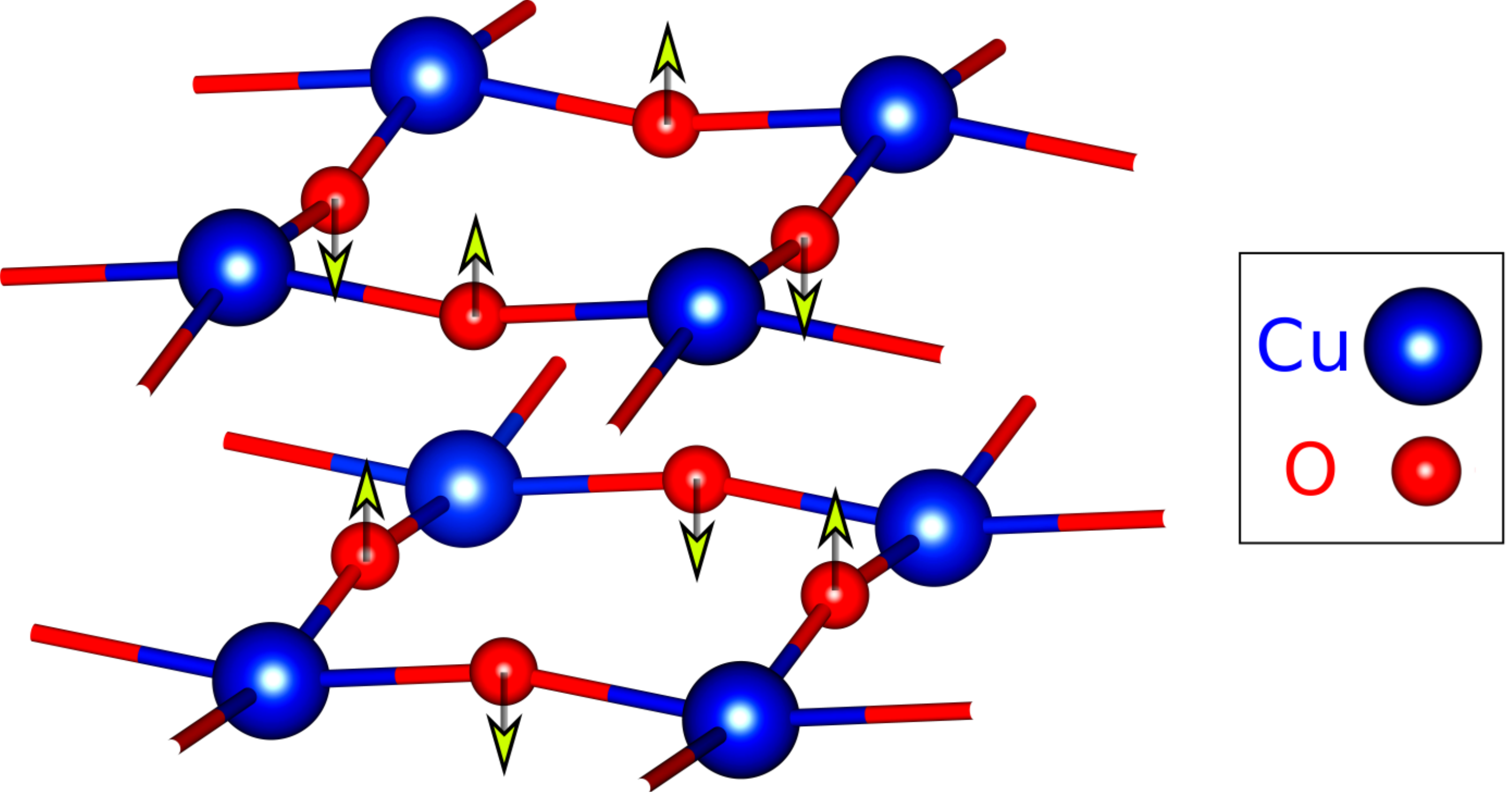} 
\caption{ A representation of the out-of-plane oxygen vibrations ($B_{1g}$ pattern) in a copper-oxide bilayer. The arrows represent the out-of-phase motions of the oxygen atoms.} \label{fig:b1gmode}
\end{figure}

 A third hint comes from the atomic displacement pattern that accompanies the CDW. There appears to be only one complete data set for the static lattice distortion pattern in charge-ordered underdoped YBCO. X-ray diffraction data in Ref.~\cite{Forgan2015} found the by far largest displacements for planar oxygen atoms; they shift out-of-plane with an out-of-phase pattern (see Fig.~\ref{fig:b1gmode}) that closely resembles the normal mode of the B$_{1g}$ bond-buckling phonon~\cite{PhysRevLett.93.117003}. In stoichiometric, overdoped YBCO the dispersion of the B$_{1g}$ phonon was monitored upon cooling~\cite{PhysRevLett.107.177004}; even though the stoichiometric composition of YBCO does not support CDW order, the B$_{1g}$ phonon frequency softens by about 6$\%$ at the charge ordering wavevector for underdoped YBCO. The x-ray diffraction data in Ref.~\cite{Forgan2015} reveal that also the heavier ions in YBCO slightly move away from the positions they take in the charge homogeneous phase in equilibrium. This naturally suggests that also the low-energy phonons are involved in the CDW formation, and indeed the softening of a low-energy optical phonon was observed by Le Tacon \textit{et al.}~\cite{LeTacon2013} and Kim \textit{et al.}~\cite{Kim1040}. While the softening of this mode indeed indicates a coupling to the CDW (as must occur in the absence of special symmetries), the small static displacements of the Ba and Y atoms observed in x-ray diffraction suggests ultimately that there is a relatively weak participation of this mode in the CDW formation. Besides small additional in-plane displacements of the oxygen atoms, the x-ray experiments~\cite{Forgan2015} clearly signal that the lattice distortions in the CDW phase have predominant B$_{1g}$ character.

Based on these empirical grounds, we select the B$_{1g}$ bond-buckling phonon exclusively, reanalyze its coupling to the electrons in the copper-oxygen planes, and pursue first the Landau free energy for a B$_{1g}$-phonon mediated CDW. We start from a microscopic model for the CuO$_2$ planes, and find that the structure of the \textit{el-ph} coupling matrix element $\mathsf{g}(\vq; \vk)$  depends strongly on the orbital content of the Fermi surface.  In particular,  the Cu-4\textit{s} orbital  is crucial; with it, $\mathsf{g}(\vq; \vk)$ is maximum at phonon momenta $\vq^*$ that are axial (rather than diagonal) and track the doping dependence of the CDW wavevector. The special role of $\vq^*$  is imperceptible in the  free energy when calculated with bare electron dispersions; however, with electronic correlations, modeled by a renormalization of the band dispersion and quasiparticle spectral weight, an axial wavevector close to $\vq^*$ emerges as the dominant wavevector in the CDW susceptibility.   The $B_{1g}$ mode is, by itself, too weak to induce true long-range order; but the inevitable presence of disorder will necessarily slow down and pin the CDW fluctuations, creating local patches of static or quasi-static charge order with finite correlation lengths. This leads us to propose a scenario, in which a phonon-based mechanism is enabled by strong electronic correlations, and the incommensurate wavevector of the concomitant charge correlations is dictated by the momentum-space structure of the \textit{el-ph} coupling matrix element.


 \section{Model}
 
To set the stage for the essential features of the electronic structure we select YBCO as the target material. The unit cell of YBCO contains a bilayer of CuO$_2$ planes. The coupling between  these two layers split the individual-layer derived bands. With respect to CDW formation it was demonstrated in earlier theoretical work~\cite{PhysRevB.90.125108} that the bilayer splitting determines primarily the relative orientation and phase of the CDWs in the two separate layers but does not have much effect on the structure of the CDW in the individual layers themselves. In YBCO, the bilayer splitting even collapses in the near-nodal region at low doping~\cite{Fournier2010}. We therefore ascribe no vital role to the bilayer structure for the CDW itself and henceforth focus on a single copper-oxygen plane.

We start by modeling a single CuO$_2$ plane in YBCO including copper $4s$ and $3d_{x^2-y^2}$ as well as oxygen $p_x$ and $p_y$ orbitals. An effective three-band model is obtained in terms of the $d$- and $p$- orbitals by downfolding the original four-band model to~\cite{ANDERSEN19951573,Atkinson2015}	
 \begin{equation}\label{Model}
\scalebox{0.9}[1]{$H_{kin} = \sum_{\vk,\sigma}\bm{\Psi^{\dagger}_{\vk,\sigma}}
\begin{pmatrix}
\varepsilon_d  & 2t_{pd}s_x    & -2t_{pd}s_y\\
    2t_{pd}s_x     & \tilde{\varepsilon}_x(\vk) & 4\tilde{t}_{pp}s_x s_y \\
    -2t_{pd}s_y    & 4\tilde{t}_{pp}s_xs_y & \tilde{\varepsilon}_y(\vk)
\end{pmatrix} \bm{\Psi}_{\vk,\sigma},$}
 \end{equation}
with the three-spinor $\bm{\Psi^{\dagger}_{\vk,\sigma}} = \left( d^{\dagger}_{\vk,\sigma},\; p^{\dagger}_{x\vk,\sigma},\; p^{\dagger}_{y \vk,\sigma}\right)$ and $s_{x,y} = \sin(k_{x,y}/2)$. $\varepsilon_d$ denotes the onsite energy of the $d$ orbital, $t_{pd}$ and $t_{ps}$ are the hopping amplitudes between $p$- and $d$- and $p$- and $s$- orbitals, respectively. In the downfolding procedure, the hopping processes via the copper $4s$ orbital renormalize the oxygen energies $\varepsilon_p$ and generate indirect hopping $4t^i_{pp}$ between oxygen orbitals:\begin{equation}\label{Onsite}
\tilde{\varepsilon}_{x,y} = \varepsilon_p + 4t^i_{pp} s^2_{x,y};\; \tilde{t}_{pp} = t^i_{pp} + t^d_{pp};\;  t^i_{pp} = \frac{t^2_{ps}}{\varepsilon_F-\varepsilon_s},
\end{equation}
where $\varepsilon_F$ is the Fermi energy and $t^d_{pp}$ a small direct hopping amplitude. We adopt all the parameters entering Eqs.~\ref{Model} and \ref{Onsite} from Ref.~\cite{ANDERSEN19951573}, specifically $t_{pd} = 1.6$ eV, $\varepsilon_d-\varepsilon_p = 0.9$ eV, $t^d_{pp} = 0$ and $t^i_{pp} = -1.0$ eV. We diagonalize  $H_{kin}$ and focus only on the partially filled anti-bonding band; the irrelevant spin index is subsequently suppressed.

\begin{figure}[t]
\centering
\includegraphics[width=1\linewidth]{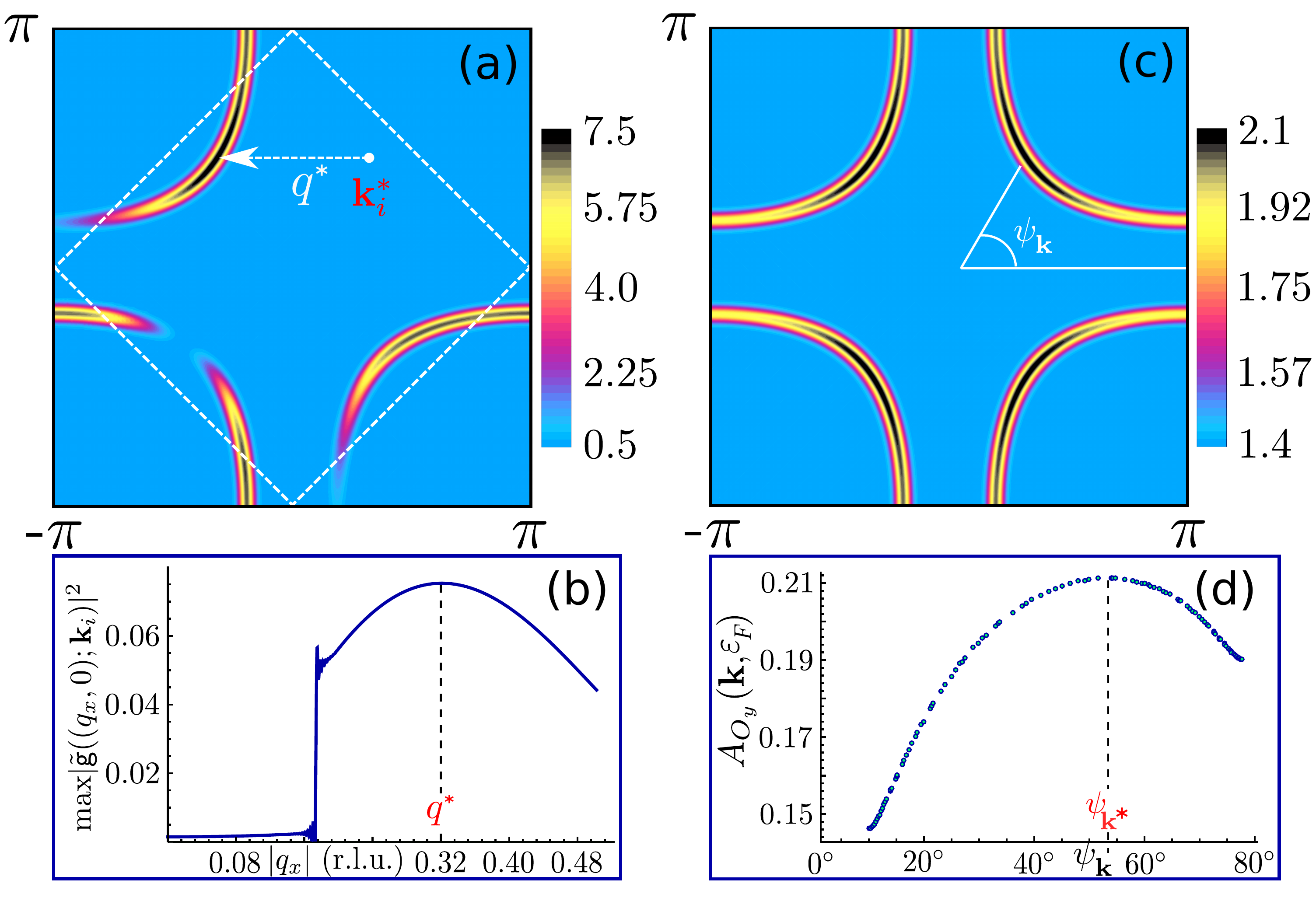}  
\caption{(a) Plot of the dimensionless \textit{el-ph} coupling $|\tilde{\mathsf{g}}(\vq;\vk^*)|^2$ (in units of $10^{-2}$) for phonon wavevectors $\vq$ that connect the \textit{initial} $\vk^*_i$ and the \textit{scattered} $\vk^*_i$$+$$\vq$ state, both on the Fermi surface at $10\%$ hole doping. The strongest coupling occurs at the axial wavevector $\vq^*$. (b) The maximum value of $|\tilde{\mathsf{g}}(\vq;\vk)|^2$ with respect to all initial momenta $\vk_i$ for axial wavevectors $\vq = (q_x,0)$. The global maximum is achieved at $\vq^*$ for the initial state at $\vk^*_i$. (c) Spectral function $A^{y}(\vk,\varepsilon_F)$ (in units of $10^{-1}$ eV$^{-1}$) for the oxygen $p_y$-orbital electron on the Fermi surface. (d) The variation of $A^{y}(\vk,\varepsilon_F)$ along the Fermi surface parametrized by the angle $\psi_{\vk}$ indicated in panel (c). } \label{fig:Result1}
\end{figure}

 The out-of-plane $B_{1g}$ vibrations of the oxygen atoms, as depicted in Fig.~\ref{fig:b1gmode}, naturally couple linearly to their local electric field $E_z$~\cite{PhysRevB.51.505}. At this point, we neglect the motion of the almost four times heavier copper atoms. Consequently, we start from the ansatz 
 \begin{equation}\label{ep-ph}
\mathcal{H}_{e\-p} = eE_z \sum_{\mathbf{n}}\Big[ \delta u_{x\mathbf{n}}  p^{\dagger}_{x \mathbf{n}}p_{x \mathbf{n}} + \delta u_{y\mathbf{n}} p^{\dagger}_{y\mathbf{n}}p_{y \mathbf{n}}\Big],
 \end{equation}
where $\delta u_{x,y \mathbf{n}}$ are the out-of-plane displacements of the oxygen atoms in unit cell $\bm{n}$. We project the \textit{el-ph} Hamiltonian in Eq.~\ref{ep-ph}, onto the anti-bonding band and obtain the effective Hamiltonian $\mathcal{H} = \mathcal{H}_{\textit{el-ph}}+ \mathcal{H}_{\textit{ph}}$ with 
\begin{equation}\label{effective_model}
\scalebox{0.9}[1]{$\mathcal{H}_{\textit{el-ph}} = \sum_{\vk} \varepsilon_{\vk} c^{\dagger}_{\vk}c_{\vk} + \sum_{\vq,\vk}\mathsf{g}(\vq;\vk)c^{\dagger}_{\vk+\vq}c_{\vk}\left( a_{\vq} + a^{\dagger}_{-\vq}\right)$},
\end{equation}
where $c^{\dagger}_{\vk}$ is the creation operator for the anti-bonding electrons with dispersion $\varepsilon_{\vk}$, $a_{\vq}$ annihilates a $B_{1g}$ phonon mode, and $\mathcal{H}_{\textit{ph}} = \hbar \Omega_P \sum_{\vq}a^{\dagger}_{\vq}a_{\vq}$ (see Ref.~\cite{PhysRevB.51.505} and Supplementary Materials). The momentum-dependent \textit{el-ph} coupling is written $\mathsf{g}(\vq;\vk) = \gamma \tilde{\mathsf{g}}(\vq;\vk)$~\cite{PhysRevB.59.14618}, where $\gamma$ is the coupling strength and 
\begin{equation}\label{coupling}
\tilde{\mathsf{g}}(\vq;\vk) =  \Big[ e^x_\vq   \phi_x(\vk')\phi_x(\vk) + e^y_\vq   \phi_y(\vk')\phi_y(\vk) \Big],
\end{equation}
with $\vk' = \vk + \vq$.  The eigenfunctions $\phi_{x,y}$ signify the orbital contents of the oxygen $p_{x,y}$ orbitals in the anti-bonding band. Similarly, the eigenvectors $e^{x,y}_{\vq}$ correspond to the normal mode of the out-of-plane displacements of the two oxygen atoms in the CuO$_2$ unit cell. The overall strength of the \textit{el-ph} coupling is $\gamma = eE_z\sqrt{\hbar/2m \Omega_P}$, where $m$ is the mass of the oxygen atom and $\Omega_P\sim40$ meV is the frequency of the dispersionless $B_{1g}$ mode. Adopting the electric-field value $eE_z=3.56$ eV/{\AA} from Ref.~\cite{PhysRevB.82.064513} leads to $\gamma = 0.22$ eV.  The eigenvectors for the $B_{1g}$ mode are $e^{x,y}_{\vq} = \mp \frac{\cos (q_{y,x}/2)}{M_{\vq}}$, where the normalization factor $M_{\vq}=\sqrt{\cos^2{(q_x/2)}+\cos^2{(q_y/2)}}$.

 Earlier theoretical work~\cite{PhysRevLett.93.117003,PhysRevLett.93.117004} on the $B_{1g}$ phonon in optimally doped Bi-2212, argued that for an antinodal fermion state, $|\mathsf{g}(\vq;\vk)|^2$ is strongest for an axial scattering wavevector $\vq$ to the nearby antinodal final state.  This value of $\vq$ is considerably smaller than $\vq_\mathit{CO}$.
 Instead we find, when the Cu-$4s$ orbital is properly included via the finite indirect hopping $t^i_{pp}$ in Eq.~\ref{Onsite}, the anisotropic structure of $\mathsf{g}(\vq;\vk)$ changes qualitatively. The maximum of the coupling $|\mathsf{g}(\vq;\vk)|^2$ now occurs for larger axial wavevectors $\vq^*$, that connect initial ($\vk^*_i$) and final ($\vk^*_i$$+$$\vq^*$) Fermi surface states near the nodal points (see Fig.~\ref{fig:Result1}a). The resulting $\vq^\ast$ is quantitatively close to experimental values of $\vq_\mathit{CO}$.

 Fixing an initial state $\vk_i$ on the Fermi surface, we evaluate the maxima of $|\mathsf{g}(\vq;\vk_i)|^2$ with respect to $\vq$ where $\vk_i+\vq$ is the final state on the Fermi surface, using the Nelder-Mead~\cite{doi:10.1137/S1052623496303470} gradient approximation. We repeat this procedure as we vary the initial state $\vk_i$ along the Fermi surface branch in the first BZ quadrant and thereby identify the global maximum which we denote as $|\mathsf{g}(\vq^*;\vk^*_i)|^2$. A plot of $|\tilde{\mathsf{g}}(\vq;\vk^*_i)|^2$ versus phonon wavevector $\vq$ is shown in Fig.~\ref{fig:Result1}a. The strongest scattering occurs at the axial $\vq^*$ indicated by the white arrow.
 
 In order to determine what is special about $\vk^*_i$ and $\vq^*$, we show the oxygen $p_y$-orbital resolved spectral function $A^{y}(\vk,\varepsilon_F)$ in Fig.~\ref{fig:Result1}c. The highest spectral weight is obtained for $\vk = \vk^*_i$. In Fig.~\ref{fig:Result1}d we parametrize the position on the Fermi surface by the angle $\psi_{\vk} = \tan^{-1}(|k_y/k_x|)$. This panel indicates that it is the oxygen orbital content on the Fermi surface that determines $\vk^*_i$ and $\vq^*$. The variation of the oxygen content is specifically controlled by the indirect hopping processes via the Cu-$4s$ orbital.
 
 \section{Analysis and Results}

\begin{figure}[t]
\centering
\includegraphics[width=1\linewidth]{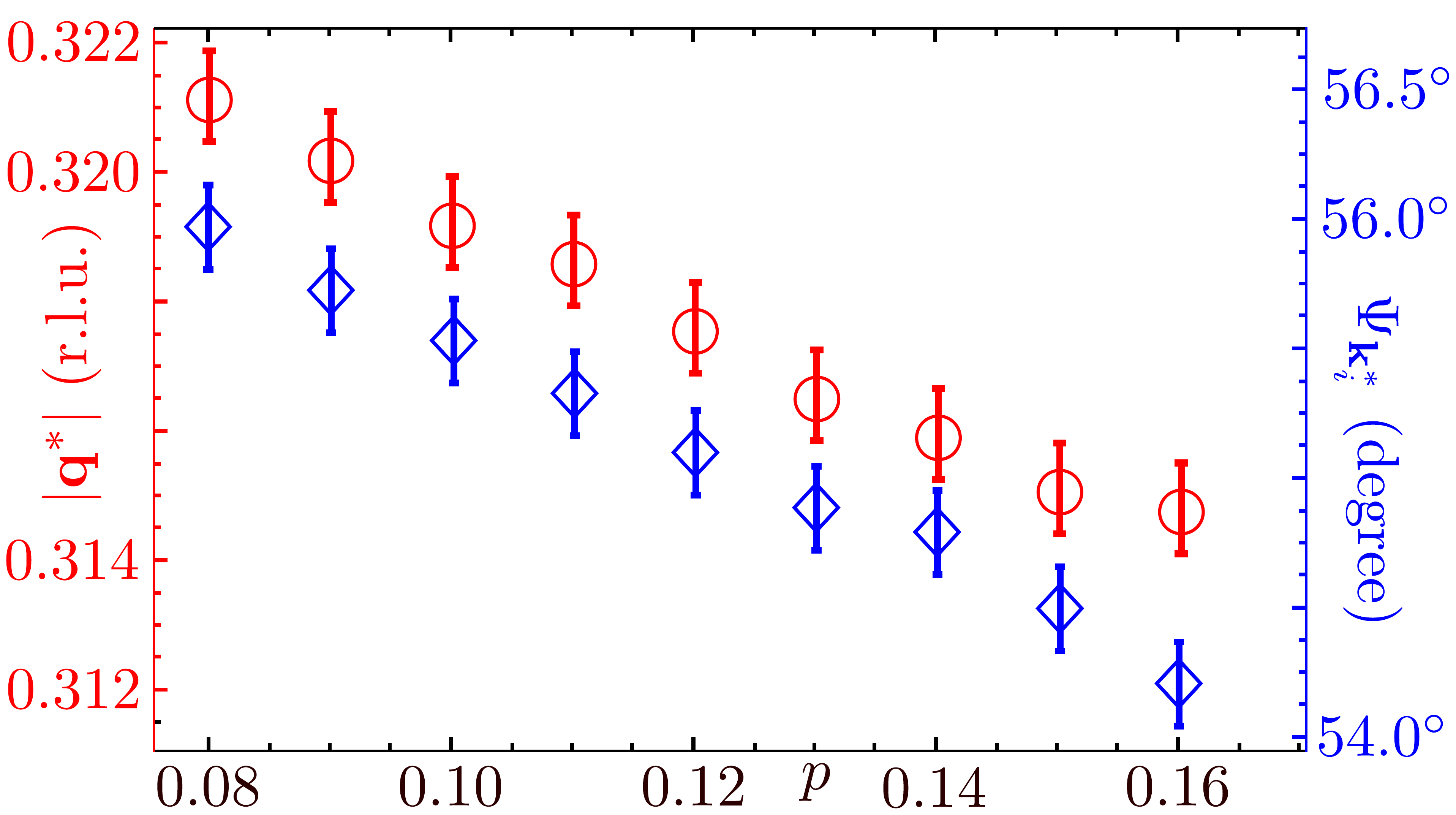}  
\caption{Hole doping dependence of the axial wavevector $\vq^*$ (\textit{circles}) and the angle $\psi_{\vk^*_i}$ (\textit{diamonds}) for the initial electronic momentum $\vk^*_i$ at which the global maximum of $|\mathsf{g}(\vq;\vk)|^2$ is achieved. The error bars represent the $\vk$-space resolution.} \label{fig:Result2}
\end{figure}

 We next calculate the doping evolution of $\vq^*$ and $\vk^*_i$ and collect the results in Fig.~\ref{fig:Result2}. The magnitude of  $\vq^*$ comes close to the observed CDW ordering wavevector $0.3$ (r.l.u.) and decreases with hole doping in a similar fashion as detected in x-ray experiments~\cite{Comin123, Comin390, NetoS393, Chang2012, PhysRevLett.110.137004, Peng2018}. It is therefore tempting to suspect a close connection. To pursue this idea, we return to the \textit{el-ph} Hamiltonian Eq.~\ref{effective_model} and assume a static mean-field lattice distortion
\begin{equation}\label{ansatz}
\sqrt{\frac{\hbar}{2m\Omega_P}} \Braket{a_{\vq} + a^{\dagger}_{-\vq}} = \xi_{\vq},
\end{equation}
with the four possible axial wavevectors $\pm \vq^*,\pm \overline{\vq}^*$ oriented either along the $x$- or the equivalent $y$- direction. We perform a linked-cluster expansion for the free energy to low orders in $\Delta_{\vq} = eE_z \xi_{\vq}$ (see  Supplementary Material and Refs.~\cite{PhysRevB.89.024507,PhysRevB.74.245126,Mahan2011}):
\begin{align}\label{free_energy}
\mathcal{F}  =  \mathcal{F}_0 + & \sum_{\vq = \vq^*, \overline{\vq}^*} \Big[\frac{|\Delta_{\vq}|^2}{4\gamma^2} \left( \hbar \Omega_P- 2\chi^{(\mathsf{g})}_{\vq}\right) +  |\Delta_{\vq}|^4 \chi^{(4)}_{\vq} \Big]  \nonumber \\
&+ |\Delta_{\vq^*}|^2 |\Delta_{\overline{\vq}^*}|^2 \chi^{(4)}_{\vq^*;\overline{\vq}^*}.
\end{align}
The static susceptibility $\chi^{(\mathsf{g})}_{\vq}$ is defined as
\begin{equation}\label{susc_charge}
\chi^{(\mathsf{g})}_{\vq} = -2\gamma^2\sum_{\vk} |\tilde{\mathsf{g}}(\vq;\vk)|^2 \Bigg[\frac{f(\varepsilon_{\vk})-f(\varepsilon_{\vk+\vq})}{\varepsilon_{\vk}-\varepsilon_{\vk+\vq}}\Bigg],
\end{equation}
where $f(\varepsilon)$ denotes the Fermi distribution function. Upon cooling from high temperature, a lattice instability occurs at the $\vq$ value for which the coefficient $(\hbar \Omega_P- 2\chi^{(\mathsf{g})}_{\vq})$ in the quadratic term first vanishes. This instability will necessarily produce an incommensurate charge modulation at the same $\vq$. (In previous theoretical work, such a criterion was successfully employed to identify the correct CDW wavevectors for weakly correlated tellurides~\cite{PhysRevB.74.245126}.)
Upon further cooling, the fourth-order coefficients decide between uniaxial and biaxial charge order.  Because the magnitudes  of the coefficients, and even the sign of $\chi^{(4)}_{\vq^*;\overline{\vq}^*}$, depend sensitively on specific parameter choices,  it is difficult to make universal statements about the behavior of the quartic terms.

\begin{figure}[t]
\centering
\includegraphics[width=0.97\linewidth]{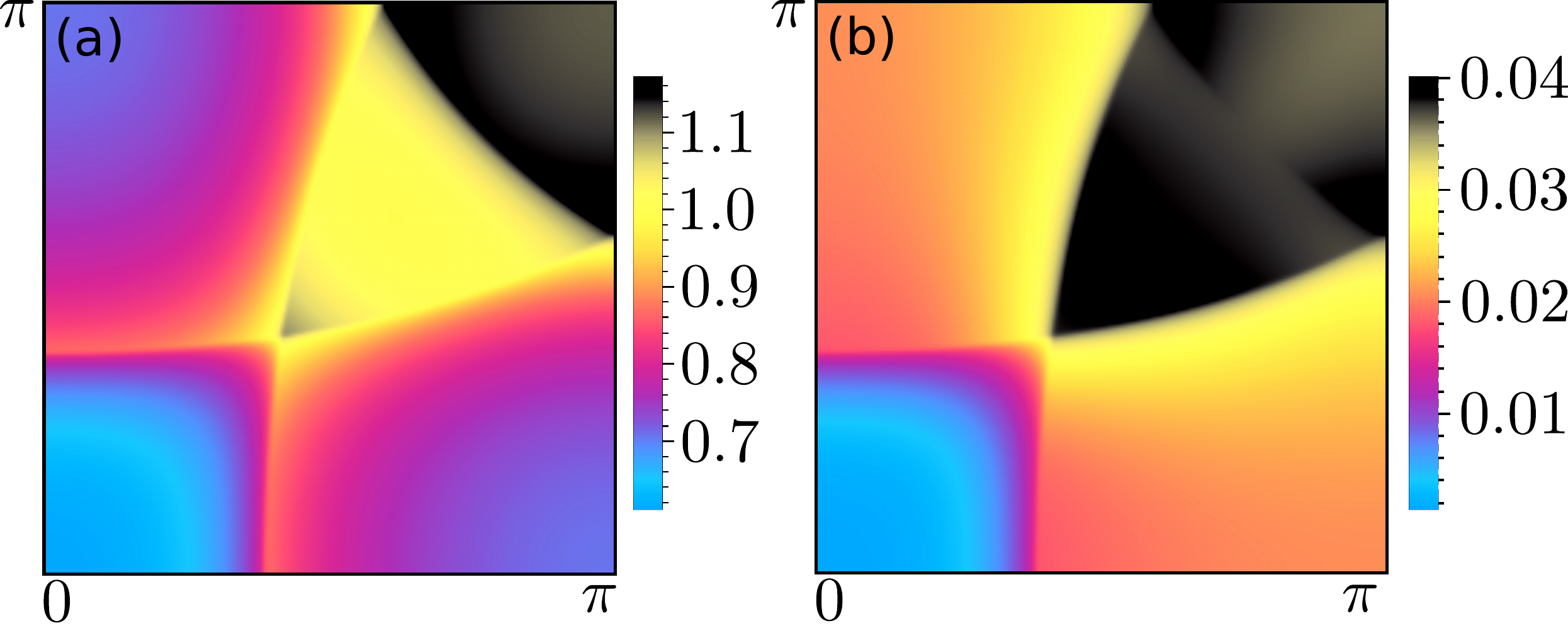}
\caption{(a) Lindhard function $\chi^{(\mathrm{L})}_{\vq}/\gamma^2$ 
and (b) $\chi^{(\mathsf{g})}_{\vq}/\gamma^2$ (in units of (eV)$^{-1}$) 
in the first quadrant of the BZ at $110$ K and $10\%$ hole doping.}\label{fig:Result4}
\end{figure}

 Returning to the quadratic term, we define for comparison the Lindhard function, $\chi^{(\mathrm{L})}_{\vq}$, obtained by setting $\tilde{\mathsf{g}}(\vq;\vk) = 1$ in Eq.~\ref{susc_charge}; as shown in Fig.~\ref{fig:Result4}(a), its weight is concentrated near the $(\pi,\pi)$ point without any prominent wavevector related to nesting. Yet, as demonstrated in Refs.~\cite{PhysRevB.77.165135,Eiter2013}, the momentum dependence of the \textit{el-ph} coupling can by itself select the wavevector for a lattice instability and the concomitant charge order. Indeed, if the structure of $\tilde{\mathsf{g}}(\vq;\vk)$ is incorporated as in Eq.~\ref{susc_charge}, the momentum-space weight of $\chi^{(\mathsf{g})}_{\vq}$ redistributes [Fig.~\ref{fig:Result4}(b)]. But, a clear instability wavevector still cannot be seen, and the largest values of $\chi_{\vq}^{(\mathsf{g})}$ near the 2 meV are an order of magnitude smaller to meet the instability condition $2\chi_{\vq}^{(\mathsf{g})} = \hbar \Omega_P = 40$ meV. In particular, the axial wavevector $\vq^*$ for the global maximum of $|\tilde{\mathsf{g}}(\vq;\vk)|^2$ remains invisible. 
 
The above analysis for the \textit{el-ph} Hamiltonian Eq.~\ref{effective_model} apparently fails to find an instability at the axial candidate wavevector $\vq^*$. This naturally compromises the attempt to carry over a strategy, which so far has relied only on the electronic input parameters from density-functional theory (DFT), to strongly correlated cuprates. Seeking for the role the B$_{1g}$ phonon in the CDW formation therefore requires to include quasiparticle renormalization effects beyond the DFT framework. Yet, including phonons in a multi-orbital model with strong electronic correlations is a demanding task. We therefore proceed with a trial ansatz to include correlation physics on a phenomenological level. 

The prevailing issue in the physics of underdoped cuprates is the conundrum of the pseudogap~\cite{RevModPhys.78.17}. The Fermi surface appears truncated to Fermi arcs centered around the BZ diagonals (the nodal regions) and it is obliterated near the BZ boundaries (the antinodal regions). In essence, well defined quasiparticles exist for near-nodal directions, while they are wiped out towards the antinodes. This motivates a simple ansatz for the Green function of the anti-bonding band quasiparticles, similar in spirit to Ref.~\cite{PhysRevB.89.024507}:
\begin{subequations}\label{Fudge}
\begin{align}
\label{Fudge:a}
& G(\vk, \omega)  = \frac{Z_{\vk}}{\omega - \varepsilon_{\vk} + i \delta} + G_{\text{incoh}},\\
\label{Fudge:b}
Z_{\vk} & = \left\{
\begin{array}{@{}rl@{}}
&\frac{1-\cos\left( \psi_{\vk} - \psi_{max} \right)}{1-\cos \left( 45^{\circ}-\psi_{max}\right)};\quad  \psi_{\vk} \geq 45^{\circ}\\
& \frac{1-\cos\left( \psi_{\vk} - \psi_{min} \right)}{1-\cos \left( 45^{\circ}-\psi_{min}\right)};\quad  \psi_{\vk} < 45^{\circ}
\end{array},
\right . 
\end{align}
\end{subequations}
for $\vk$ and $\psi_{\vk}$ defined in the first quadrant of the BZ (see Fig.~\ref{fig:Result1}c). $Z_{\vk}$ is the quasiparticle weight tailored to continuously decrease from 1 at the nodal point to zero at the BZ faces. $\psi_{max/min}$ denotes the largest/smallest angle $\psi_{\vk}$ for the Fermi surface momenta $(k_F,\pi)$ and $(\pi,k_F)$, respectively. The $\vk$-dependence in Eq.~\ref{Fudge:b} is analogously carried over to the other segments of the Fermi surfaces. We emphasize that Eq.~\ref{Fudge:b} is an ad hoc, phenomenological ansatz. The desired quantitative information \--- in particular with sufficient momentum-space resolution \--- for the anisotropic quasiparticle-weight renormalization at different doping levels is not available. The latter is a hard task by itself for strongly correlated electron theory.
 
 The ansatz in Eq.~\ref{Fudge:a}, \ref{Fudge:b} leads to the renormalized static susceptibility
\begin{equation}\label{renorm_susc}
\scalebox{0.9}[1]{$\chi^{(\mathsf{g}Z)}_{\vq} \approx -2\gamma^2 \sum_{\vk} Z_{\vk}Z_{\vk+\vq}|\tilde{\mathsf{g}}(\vq;\vk)|^2 $} \Bigg[\frac{f(\varepsilon_{\vk})-f(\varepsilon_{\vk+\vq})}{\varepsilon_{\vk}-\varepsilon_{\vk+\vq}}\Bigg],
\end{equation}
where the contributions from the incoherent part of $G(\vk,\omega)$ are neglected (see Supplementary Material).

\begin{figure}[t]
\centering
\includegraphics[width=1\linewidth]{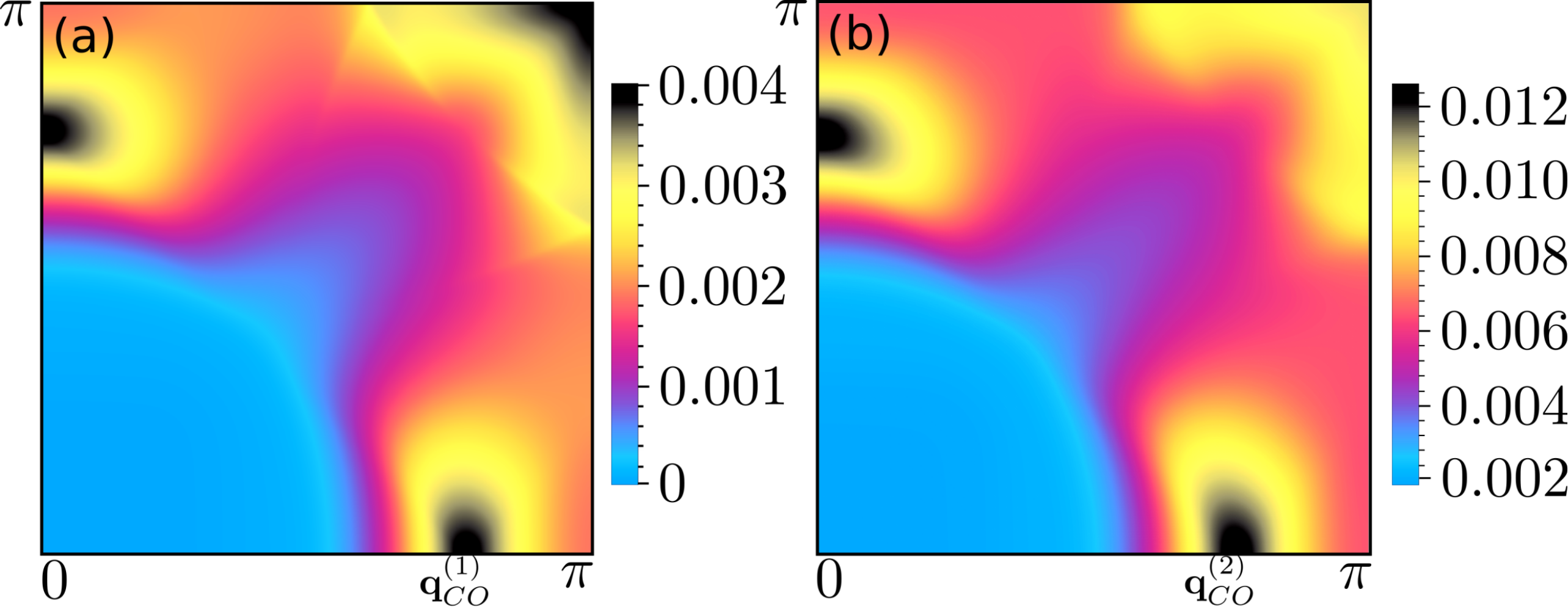}
\caption{(a) Renormalized static susceptibility $\chi^{(\mathsf{g}Z)}_{\vq}/\gamma^2$ and (b) $\chi^{(\mathsf{g}Z)}_{\vq}/\gamma^2$ (in units of (eV)$^{-1}$) modified further by a renormalized dispersion in the square brackets of Eq.~\ref{renorm_susc}, in the first quadrant of the BZ at $110$ K and $10\%$ hole doping.}\label{fig:Result5}
\end{figure}

We examine the impact of the quasiparticle weight factors on the susceptibility $\chi^{(\mathsf{g}Z)}_{\vq}$ in Fig.~\ref{fig:Result5}a. The highly anisotropic variation of $Z_{\vk}$ reduces the susceptibility around the BZ diagonal and creates new structures along the axes including the global maximum at axial wavevectors $\vq^{(1)}_{CO}$, and $\overline{\vq}^{(1)}_{CO}$, which are larger than, but close to, $\vq^*$ or $\overline{\vq}^*$ for which the \textit{el-ph} coupling is strongest.
 
In a second step we incorporate \--- besides the anisotropic weight factors $Z_{\vk}$ \--- also the correlation induced renormalization of the band dispersion by replacing $\varepsilon_{\vk}\Rightarrow \varepsilon^{R}_{\vk}$ in Eq.~\ref{Fudge:a} and hence also in the square bracket of Eq.~\ref{renorm_susc}. For this purpose, we adopt a phenomenological fit to the measured ARPES dispersion applied previously to data for optimally doped Bi-2212~\cite{PhysRevB.75.184514}. Compared to the bare dispersion, $\varepsilon^{R}_{\vk}$ has an almost three times narrower bandwidth and the nodal Fermi velocity $v_F$ is similarly reduced, while the shape of the Fermi surface remains almost preserved. The susceptibility $\chi^{(\mathsf{g}Z)}_{\vq}$ with the renormalized $\varepsilon^{R}_{\vk}$ is shown in Fig.~\ref{fig:Result5}b. We observe that the axial peaks are retained at wavevectors $\vq^{(2)}_{CO}$ with $q^*<q^{(2)}_{CO}<q^{(1)}_{CO}$, whereas the peak at the $(\pi,\pi)$ point loses its strength.  $q^{(2)}_{CO}$ is about $16\%$ larger than $q^*$ and follows the same doping dependence as $q^*$.  Furthermore, the strength of the peak at $\vq^{(2)}_{CO}$ has tripled. This increase is naturally tied to the downward renormalization of $v_F$. So,   we conclude that it apparently requires both a correlation-induced quasiparticle renormalization, and a specific momentum dependence of the \textit{el-ph} coupling to generate the axial and incommensurate candidate wavevector $\vq^{(2)}_{CO}$ for the anticipated lattice and  concomitant CDW instability. 

In reaching our conclusion, we have manifestly neglected the life-time broadening effects on the nodal quasiparticles; material difference matter. While the majority of ARPES experiments, based on Bi-based cuprates~\cite{Li2018}, report a quantitatively larger quasiparticle broadening ($\sim 10$ meV near the gap node~\cite{PhysRevB.75.140513}), state-of-the-art transport experiments provide much smaller broadening in YBCO.  Notably, microwave measurements in YBa$_2$Cu$_3$O$_{6.5}$ find transport scattering rates of order 0.1 meV~\cite{PhysRevB.74.104508}, which is 100 times smaller than typical values quoted for Bi-2212, which is most likely connected to the inhomogeneity of the samples. Indeed, STM experiments~\cite{Pan2001,RevModPhys.79.353} show that Bi based cuprates (Bi-2201 and Bi-2212) host a relatively large material inhomogeneity, while YBCO is believed to be homogeneous. It seems to us likely that scattering rates obtained from ARPES in Bi-2212 include significant disorder broadening at low temperature and energy, and are not relevant to YBCO.

Without aiming at a quantitative description we nevertheless translate the obtained result into an estimate. The absolute units in Fig.~\ref{fig:Result5}b indicate that $\chi^{\mathsf{g}Z}_{\vq^{(2)}_{CO}} \ll \hbar\Omega_P$, and we are therefore far from meeting the mean-field instability criterion. Nonetheless, the \textit{el-ph} coupling inevitably also alters the frequency of the participating phonons. Based on Ref.~\cite{Gruner}, the renormalized $B_{1g}$ phonon frequency $\tilde{\omega}_{\vq}$ follows from 
\begin{equation}\label{Gruner}
\tilde{\omega}^2_{\vq} = \Omega^2_P(1-\lambda_{\vq}),\quad \lambda_{\vq} =
\frac{2\chi^{(\mathsf{g}Z)}_{\vq,\Omega_P}}{\hbar\Omega_P},
\end{equation}  
where $\chi^{(\mathsf{g}Z)}_{\vq,\Omega_P}$ is the real part of the susceptibility at the bare phonon frequency $\Omega_P$. At $110K$ and $\vq = \vq^{(2)}_{CO}$,  the dynamical susceptibility is slightly smaller than its static zero-frequency value, from which Eq.~\ref{Gruner} gives $\tilde{\omega}_{\vq} \simeq 0.987\Omega_P$, i.e. a softening of 1.3\%. Upon cooling $\chi^{(\mathsf{g}Z)}_{\vq,\Omega_P}$  increases slightly and the softening reaches $\sim$ 1.5\% at around $3K$. For reference we mention that in Ref.~\cite{PhysRevLett.107.177004}, the measured softening of the B$_{1g}$ phonon frequency in YBa$_2$Cu$_3$O$_7$ was about 6$\%$ at $3K$ for $\vq \sim (0,0.3)$.   

Several factors may act to enhance the \textit{el-ph} coupling in underdoped cuprates. We recall that a dispersion $\varepsilon^R_{\vk}$ for an optimally doped material was used for the evaluation of $\chi^{(\mathsf{g}Z)}_{\vq}$. On the underdoped side, the nodal Fermi velocity in Bi-2212 drops by as much as 50\%~\cite{PhysRevLett.104.207002}. Such a drop necessarily enhances $\chi^{(\mathsf{g}Z)}_{\vq^{(2)}_{CO}}$ and the corresponding phonon softening is estimated to rise to about $2.12\%$ reflecting somewhat enhanced CDW correlations at wavevectors $\vq^{(2)}_{CO}$ for underdoped materials. Furthermore, strong correlations in the underdoped cuprates decrease the copper and increase the oxygen orbital content on the Fermi surface, thereby enlarging the susceptibility at $\vq^{(2)}_{CO}$. One may argue that a considerably larger charge susceptibility \--- as an outcome of our calculation \--- would strengthen the case for the proposed mechanism. Yet, any theory for the CDW in Y- or Bi-based cuprates has to be reconciled with the prevailing fact that ARPES experiments fail to identify signatures of spectral changes or electronic reconstruction in the charge ordered phase. Even, in charge-stripe ordered Nd-doped LSCO, the interpretation of line-shape changes remains subtle~\cite{PhysRevB.92.134524,Kivelson14395}. For these reasons, the anticipated CDW can be considered weak.

The rough estimates presented above for an only weak phonon-based tendency towards charge order is \--- in this sense \--- not contradicting, but it is clearly too weak to generate a long range ordered CDW state. Still, the charge order in underdoped cuprates is in fact short ranged with moderate planar correlation lengths~\cite{Ghiringhelli821,Chang2012,PhysRevLett.110.137004},  and is most likely nucleated by defects~\cite{Campi2015,Wu2015,PhysRevB.97.125147}. A strong magnetic field~\cite{Gerber949,Jang14645} or uniaxial pressure~\cite{Kim1040} is needed to enhance the planar correlation length and to even achieve 3D CDW order. The magnetic field suppresses superconductivity, which otherwise stops the charge correlations from growing upon cooling.

\section{Discussion and Concluding Remarks}

Our proposed mechanism therefore appears compatible with experimental observations. The momentum-dependence of the \textit{el-ph} coupling to the $B_{1g}$ bond-buckling phonon and the specific variation of the oxygen orbital content on the Fermi surface select an incommensurate, axial wavevector $\vq^*$ for which the lattice is most susceptible to deform. But only in conjunction with the strong and anisotropic renormalization of the correlated electrons in the copper oxygen planes does the corresponding susceptibility develop the required enhancement near $\vq^*$ to move the \textit{el-ph} systems at least towards a charge ordered state with an axial wavevector $\vq^{(2)}_{CO}$ near $\vq^*$. 

Compared to bilayer materials like YBCO the case of single-layer cuprates is more complicated. In pristine Hg-1201 and Bi-2201, the CuO$_2$ planes have mirror symmetries that prohibit a linear coupling between the B$_{1g}$ phonon and the electrons. However, both of these materials are doped by large concentrations of interstitial oxygen atoms that reside above the CuO$_2$ plane. These unscreened dopants are the source of large electric fields in the CuO$_2$ plane and act as nucleation sites for CDW patches~\cite{PhysRevB.96.134510}. In bilayer cuprates viz. Bi-2212, such interstitial oxygen atoms generate an electric field of the order of few eV/{\AA} which in turn strongly amplifies (up to a factor of 5) the strength of the B$_{1g}$ el-ph coupling~\cite{Johnston_2009}. These field strengths are comparable to that obtained for YBCO, and must (by symmetry) produce a linear coupling between the B$_{1g}$ phonon and the CDW. Based on these empirical facts, we expect a similar strong enhancement of the B$_{1g}$ coupling in the single-layer cuprates, such as Bi-2201 and Hg-1201.

An important question is the extent to which such inhomogeneous el-ph coupling is visible in the phonon dispersion.  Provided the phonons are harmonic, the lattice distortion associated with the local electric fields should not shift the phonon frequencies; rather, the \textit{el-ph} matrix element will be distributed across a range of values leading to a broadening of the phonon dispersion near the ordering wavevector.  However, the scale of the broadening, and in particular, how it compares with other sources of apparent broadening, as discussed in Ref.~\cite{PhysRevB.101.184508}, remains unsettled.

A determination of boundaries for the CDW phase in the temperature vs. doping  phase diagram is beyond the scope of our current work, but we offer arguments for the relevant ingredients which determine the variation with respect to hole doping. The magnitude of the charge susceptibility at the anticipated ordering wavevector is controlled by i) the magnitude of the \textit{el-ph} coupling for initial and scattered electron momenta on the Fermi surface, ii) the nodal Fermi velocity, and iii) the quasiparticle weight at those Fermi surface points where $\mathsf{g}(\vq;\vk)$ achieves its largest values in the near nodal regions.

First, the global maximum value of $\mathsf{g}(\vq;\vk)$ on the doping dependent Fermi surfaces drops with hole doping, \textit{i.e.} upon leaving the underdoped region towards the overdoped side (see Sec.~E in Supplemental Material). Simultaneously, the nodal Fermi velocity almost doubles~\cite{PhysRevLett.104.207002}. Both of these trends reduce the charge susceptibility. 

The ansatz for the quasiparticle weight along the Fermi surface reflects the Fermi arc formation in underdoped cuprates; the quasiparticles are assumed intact in the near-nodal region with only a weak quasiparticle weight reduction. Necessarily we have to expect that the quasiparticle-weight - even in the near-nodal regions - will shrink upon lowering the hole doping concentration towards the insulator. Taken together, these trends naturally suggest a dome shaped CDW region in the temperature vs. doping phase diagram centered around an underdoped composition. This qualitative reasoning complies with the experimental findings.

For the intra-unit cell symmetry of the charge redistribution in the CDW state, our proposed scenario leads to a predominant $s$-symmetry form factor. Since, Cu-$d$-orbital character is admixed to the near-nodal Fermi surface points connected by $\vq^*$, also the charge on the Cu ion will be sizably modulated, as indeed is detected by nuclear magnetic resonance~\cite{Wu2011}. Although a prominent $d$-wave character for the charge modulations on the oxygen $p$-orbitals has been reported in early resonant x-ray scattering~\cite{Comin2015} and STM experiments~\cite{FujitaE3026}, this initial conclusion has recently been disputed. Instead,  the x-ray data in Ref.~\cite{2019arXiv190412929M} rather support a dominant s-wave form factor and are therefore compatible with the prediction from the phonon scenario.

At the core of our proposal is that it requires both, correlated electron physics and the coupling to the lattice degrees of freedom to address the CDW in cuprates. The special momentum-space structure of the \textit{el-ph} coupling matrix element for the B$_{1g}$ bond-buckling phonon has revealed an important ingredient which was not appreciated in electronic theories before. This is the variation of the oxygen orbital content on the Fermi surface, which dictates for which momenta the coupling, here to the B$_{1g}$ phonon, is strongest. This may prove as a relevant step forward to elucidate the true complexity of the CDW phenomenon in cuprates. Strong electronic interactions by themselves develop charge correlations, but we infer that these may lock into an incommensurate charge-ordering pattern in Y- and Bi-based cuprates only in conjunction with a specific momentum-dependent microscopic coupling to the lattice.

\section{Acknowledgement}
  
We acknowledge helpful conversations with T. P. Devereaux, P. J. Hirschfeld, T. Kopp, L. Chioncel, S. Johnston, and C. Morice. WAA acknowledges support by the Natural Sciences and Engineering Research Council (NSERC) of Canada. APK acknowledges support by the Deutsche Forschungsgemeinschaft (DFG, German Research Foundation)- project-ID-107745057-TRR 80.

\bibliographystyle{apsrev4-1}
\bibliography{Charge_cuprates}

\end{document}


\subsection*{\large Supplementary Material: \--- \\ Emergent Charge Order from Correlated Electron-Phonon Physics in Cuprates}

\begin{center}
S. Banerjee$^{1}$, W. A. Atkinson$^{2}$, and A. P. Kampf$^{1}$ \\

\small{$^1$ \textit{Theoretical Physics III, Center for Electronic Correlations and Magnetism, \\ Institute of Physics, University of Augsburg, 86135 Augsburg, Germany}} \\ \small{ $^2$ \textit{Department of Physics and Astronomy, Trent University, Peterborough, Ontario K9L 0G2, Canada}}
\end{center}

\subsection{Anti-bonding Band: Dispersion and eigenfunctions}

Diagonalizing the downfolded Hamiltonian $H_{kin}$ (Eq.~(1) in the main text), we obtain energy for the anti-bonding band as
\begin{equation}\label{anti-bonding}
\varepsilon_\vk  = \frac{\varepsilon_d+\tilde{\varepsilon}_{x}+\tilde{\varepsilon}_{y}}{3} + \left(A_{\vk} + \sqrt{A_{\vk}^2+B_{\vk}^3} \right)^{\frac{1}{3}} + \left(A_{\vk} - \sqrt{A_{\vk}^2+B_{\vk}^3} \right)^{\frac{1}{3}},
\end{equation}
where the parameters $A_{\vk}$ \& $B_{\vk}$ are defined as
\begin{align}\label{parameter}
A_{\vk} & = \frac{\varepsilon_d}{6}\left( t_x^2 +t_y^2 -2t'^2\right) +\frac{\tilde{\varepsilon}_x}{6}\left( t'^2 +t_x^2 -2t_y^2\right) +\frac{\tilde{\varepsilon}_y}{6}\left( t'^2 +t_y^2 -2t_x^2\right) - t't_xt_y\\ \nonumber
&  + \frac{\varepsilon_d^3+\tilde{\varepsilon}_x^3+\tilde{\varepsilon}_y^3}{27}-\frac{\varepsilon_d^2\tilde{\varepsilon}_x+\tilde{\varepsilon}_x^2\varepsilon_d+\tilde{\varepsilon}_x^2\tilde{\varepsilon}_y+\tilde{\varepsilon}_y^2\tilde{\varepsilon}_x+\varepsilon_d^2\tilde{\varepsilon}_y+\tilde{\varepsilon}_y^2\varepsilon_d}{18} + \frac{2\varepsilon_d\tilde{\varepsilon}_x\tilde{\varepsilon}_y}{9} \\ \nonumber
B_{\vk} &= -\frac{\varepsilon_d^2+\tilde{\varepsilon}_x^2+\tilde{\varepsilon}_y^2-\varepsilon_d\tilde{\varepsilon}_x-\tilde{\varepsilon}_x\tilde{\varepsilon}_y-\tilde{\varepsilon}_y\varepsilon_d}{9}-\frac{1}{3} \left( t_x^2 +t_y^2 + t'^2 \right),
\end{align}
where $t_x=2t_{pd}s_x$, $t_y = 2t_{pd}s_y$ and $t'=4\tilde{t}_{pp}s_xs_y$. The projections of $d$, $p_x$ and $p_y$ orbitals onto the anti-bonding band follow from the eigenfunction for the anti-bonding band as 
\begin{align}\label{eigenfunction}
\phi_d(\vk)  & = \frac{1}{N_\vk} \Big[ (\varepsilon_\vk-\tilde{\varepsilon}_x)(\varepsilon_\vk-\tilde{\varepsilon}_y) - t'^2 \Big], \; \phi_x(\vk) = \frac{1}{N_\vk} \Big[ (\varepsilon_\vk-\tilde{\varepsilon}_y) t_x - t't_y \Big], \; \phi_y(\vk)   = -\frac{1}{N_\vk} \Big[ (\varepsilon_\vk-\tilde{\varepsilon}_x) t_y - t't_x \Big]\\ \nonumber
& \qquad N_\vk  = \sqrt{\Big[ (\varepsilon_\vk-\tilde{\varepsilon}_x)(\varepsilon_\vk-\tilde{\varepsilon}_y) - t'^2 \Big]^2 + \Big[ (\varepsilon_\vk-\tilde{\varepsilon}_y) t_x - t't_y \Big]^2 + \Big[ (\varepsilon_\vk-\tilde{\varepsilon}_x) t_y - t't_x \Big]^2}.
\end{align}
We notice that $(A_{\vk}^2+B_{\vk}^3)$ is negative for all $\vk$ in the Brillouin zone. Hence, we can write Eq.~\ref{anti-bonding} as
\begin{equation}\label{anti-bonding_Extra}
\varepsilon_\vk  = \frac{\varepsilon_d+\tilde{\varepsilon}_{x}+\tilde{\varepsilon}_{y}}{3} + 2\Re \Big[\left(A_{\vk} + \sqrt{A_{\vk}^2+B_{\vk}^3} \right)^{\frac{1}{3}} \Big],
\end{equation}
therefore the anti-bonding band energy is real. A simplified version of Eq.~(\ref{eigenfunction}) and Eq.~(\ref{anti-bonding_Extra}) was obtained in Refs.~\cite{PhysRevB.59.14618,PhysRevLett.93.117004}. We show the results for the most general three-band Hamiltonian model.

\subsection{Lattice dynamical model: $B_{1g}$ oxygen vibration}

\begin{figure}[t]
\centering
\includegraphics[width=0.39\linewidth]{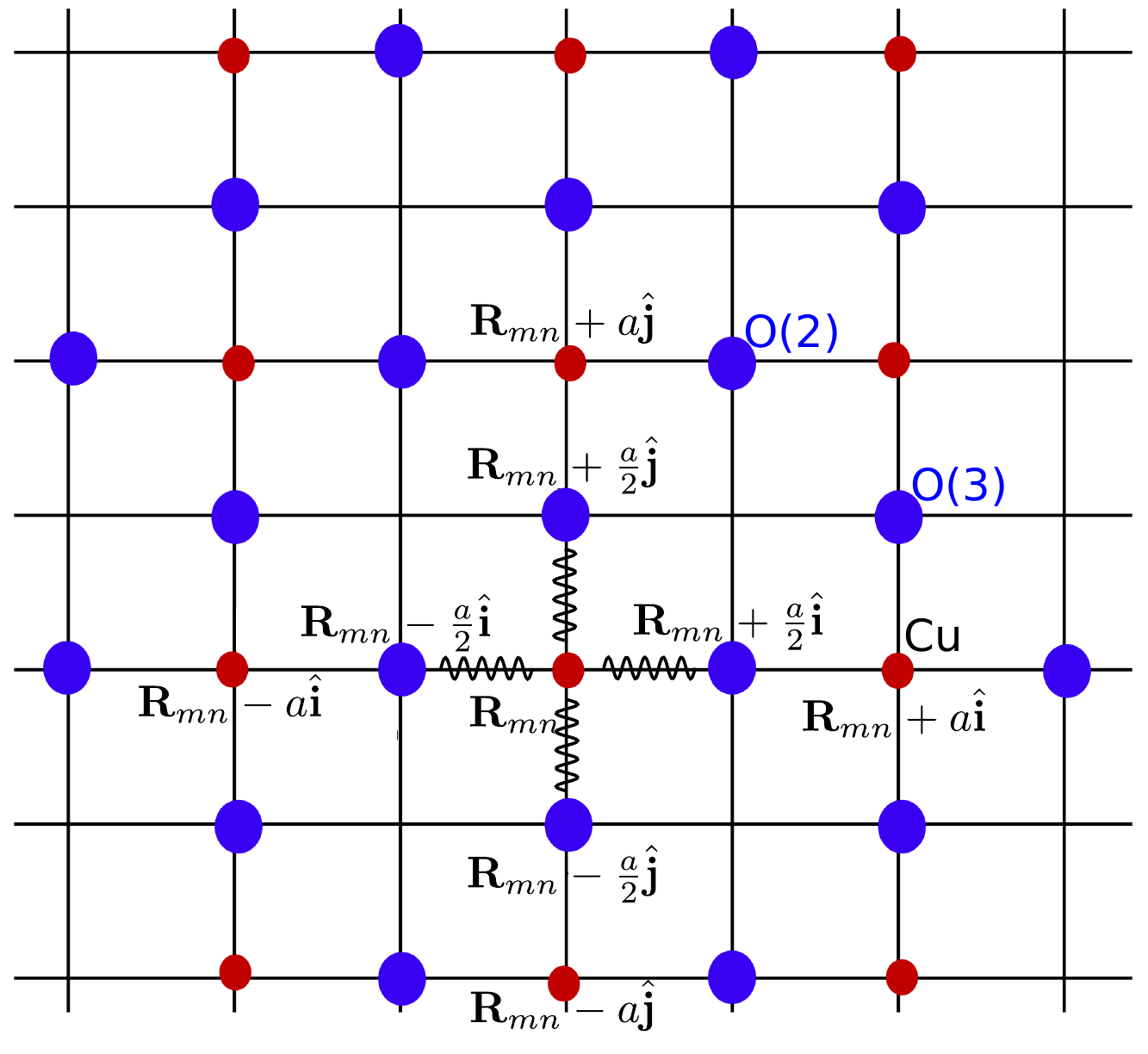}
\caption{(a) The three-atom unit cell with two oxygen atoms O(2) and O(3) coupled to copper via a uniform force constant $K$. The out-of-plane motions of the oxygen atoms are constrained by a uniform tension with force constant $K'$.}\label{fig:lattice}
\end{figure}

In this section, we analyze the eigenmodes for the out-of-plane $B_{1g}$ oxygen vibrations in a CuO$_2$ plane. We assume a three-atom unit cell of a copper and two oxygen atoms O(2,3). The B$_{1g}$ dispersion can be modeled by two spring constants: the first, $K$, is between O atoms and nearest-neighbour copper sites (Fig.~\ref{fig:lattice}); the second, $K^\prime$, represents a dispersionless coupling between the O atoms and the rest of the lattice.
Labeling the positions of Cu, O(2) and O(3) by $\vu_1(\mathbf{R})$, $\vu_2(\mathbf{R})$ and $\vu_3(\mathbf{R})$, we write the coupled equations of atomic motion as ($a$ is the lattice constant)
\begin{align}\label{one}
M\ddot{\vu}_1(\mathbf{R}_{mn}) & = -K \left(\vu_1(\mathbf{R}_{mn})-\vu_2(\mathbf{R}_{mn}+\frac{a}{2}\hat{\bm{i}})\right) - K \left(\vu_1(\mathbf{R}_{mn})-\vu_2(\mathbf{R}_{mn}-\frac{a}{2}\hat{\bm{i}})\right)\\ \nonumber
& - K \left(\vu_1(\mathbf{R}_{mn})-\vu_2(\mathbf{R}_{mn}+\frac{a}{2}\hat{\bm{j}})\right) - K \left(\vu_1(\mathbf{R}_{mn})-\vu_2(\mathbf{R}_{mn}-\frac{a}{2}\hat{\bm{j}})\right)  \\ \nonumber
m\ddot{\vu}_2(\mathbf{R}_{mn}+\frac{a}{2}\hat{\bm{i}}) & = -K \left(\vu_2(\mathbf{R}_{mn}+\frac{a}{2}\hat{\bm{i}})-\vu_1(\mathbf{R}_{mn})\right) -K \left(\vu_2(\mathbf{R}_{mn}+\frac{a}{2}\hat{\bm{i}})-\vu_1(\mathbf{R}_{mn}+a\hat{\bm{i}})\right) - K' u^z_2(\mathbf{R}_{mn}+\frac{a}{2}\hat{\bm{i}})\hat{\bm{k}} \\ \nonumber
m\ddot{\vu}_3(\mathbf{R}_{mn}+\frac{a}{2}\hat{\bm{j}}) & = -K \left(\vu_3(\mathbf{R}_{mn}+\frac{a}{2}\hat{\bm{j}})-\vu_1(\mathbf{R}_{mn})\right)-K \left(\vu_3(\mathbf{R}_{mn}+\frac{a}{2}\hat{\bm{j}})-\vu_1(\mathbf{R}_{mn}+a\hat{\bm{j}})\right) - K' u^z_3(\mathbf{R}_{mn}+\frac{a}{2}\hat{\bm{i}})\hat{\bm{k}}
\end{align}
where $M$ and $m$ are the masses of copper and oxygen atoms respectively.  
\begin{align}
\begin{bmatrix}
\frac{4K}{M} & -\frac{2K}{\sqrt{Mm}} \cos(\frac{q_x}{2}) & -\frac{2K}{\sqrt{Mm}} \cos(\frac{q_y}{2}) \\
-\frac{2K}{\sqrt{Mm}} \cos(\frac{q_x}{2}) & \frac{2K+K'}{m} & 0 \\
-\frac{2K}{\sqrt{Mm}} \cos(\frac{q_y}{2}) & 0 & \frac{2K+K'}{m}
\end{bmatrix} &
\begin{bmatrix}
\vu_{1\vq}^z\\
\vu_{2\vq}^z\\
\vu_{3\vq}^z
\end{bmatrix} = \omega^2  
\begin{bmatrix}
\vu_{1\vq}^z\\
\vu_{2\vq}^z\\
\vu_{3\vq}^z
\end{bmatrix}.
\end{align}
\begin{align}\label{eigensystem}
& \qquad \; \omega_1^2(\vq) = \frac{2K}{m}; \quad \omega_2^2(\vq)  = 0; \quad \omega_3^2(\vq) = \frac{2K}{m} \\ \nonumber
& \eta_1 = \begin{bmatrix}
0 \\ 
-\cos(\frac{q_y}{2}) \\
\cos(\frac{q_x}{2})
\end{bmatrix}, \; \eta_2 = \begin{bmatrix}
C_\vq\\ 
\cos(\frac{q_x}{2}) \\
\cos(\frac{q_y}{2})
\end{bmatrix}, \; \eta_3 = \begin{bmatrix}
-C_\vq \\
\cos(\frac{q_x}{2}) \\
\cos(\frac{q_y}{2})
\end{bmatrix}
\end{align}
where the variable $C_{\vq}$ is defined as
\begin{equation}\label{constant}
C_\vq  = \frac{\Big[2K\sqrt{mM}\left(M-2m\right) +\sqrt{\{16K^2m^3M+4K^2M^3m + 8K^2m^2M^2 (\cos q_x+\cos q_y)\}} \Big]}{4KMm}.
\end{equation}
 We now notice that $M = 63.546 u$ and $m = 15.9 u$, and thus make the simplifying assumption $M \gg m$.
In this limit, the normalized $B_{1g}$ eigendisplacement from Eq.~(\ref{eigensystem}) is 
\begin{equation}\label{eigendisp}
e^x_{\vq}  = - \hat{\bm{z}}\frac{\cos(\frac{q_y}{2})}{\sqrt{\cos^2(\frac{q_x}{2}) + \cos^2(\frac{q_y}{2})}}; \quad e^y_{\vq}   =  \hat{\bm{z}} \frac{\cos(\frac{q_x}{2})}{\sqrt{\cos^2(\frac{q_x}{2}) + \cos^2(\frac{q_y}{2})}}.
\end{equation}
Our dynamical phonon model is specifically tailored to obtain the eigendisplacements for the $B_{1g}$ vibrations. We finally, provide the variation of the global maximum of the momentum-dependent \textit{el-ph} coupling $|\tilde{\mathsf{g}}(\vq;\vk)|^2$ in Fig.~\ref{fig:doping}

\subsection{Linked cluster Expansion: Free energy}

 In this section, we outline the derivation of the mean-field Landau Free energy (Eq.~(7) in the main text). The mean-field \textit{el-ph} Hamiltonian for the static mean-field lattice distortion $\xi_{\vq}$ follows from (Eq.~(4) in the main text) as
\begin{equation}\label{Mean_field}
H_{MF} = \underbrace{\sum_{\vk} \varepsilon_{\vk} c^{\dagger}_{\vk} c_{\vk}}_{H_0} + \underbrace{\sum_{\vk} \sum_{\vq=\pm \vq^*,\pm \overline{\vq}^*} \gamma \sqrt{\frac{2m\Omega_P}{\hbar}}\mathsf{\tilde{g}}(\vq;\vk) \xi_{\vq} c^{\dagger}_{\vk+\vq}c_{\vk}}_{H_{\mathrm{EP}}} + \underbrace{\frac{m\Omega^2_P}{2}\sum_{\vq=\pm \vq^*,\pm \overline{\vq}^*} |\xi_{\vq}|^2}_{H_{\mathrm{Ph}}}.
\end{equation}  
Following Eq.~(\ref{Mean_field}), the mean-field $\xi_{\vq}$ is evaluated by minimizing the Landau free energy of the coupled \textit{el-ph} system. Utilizing the linked cluster theorem, the latter is obtained as
\begin{equation}\label{Free_energy}
\mathcal{F} = \mathcal{F}_0 + \frac{1}{2}\sum_{\vq= \vq^*,\overline{\vq}^*} m\Omega_P^2 |\xi_{\vq}|^2 - \sum_{l=1}^{\infty} U_l,
\end{equation}
\begin{figure}[b]
\centering
\includegraphics[width=0.59\linewidth]{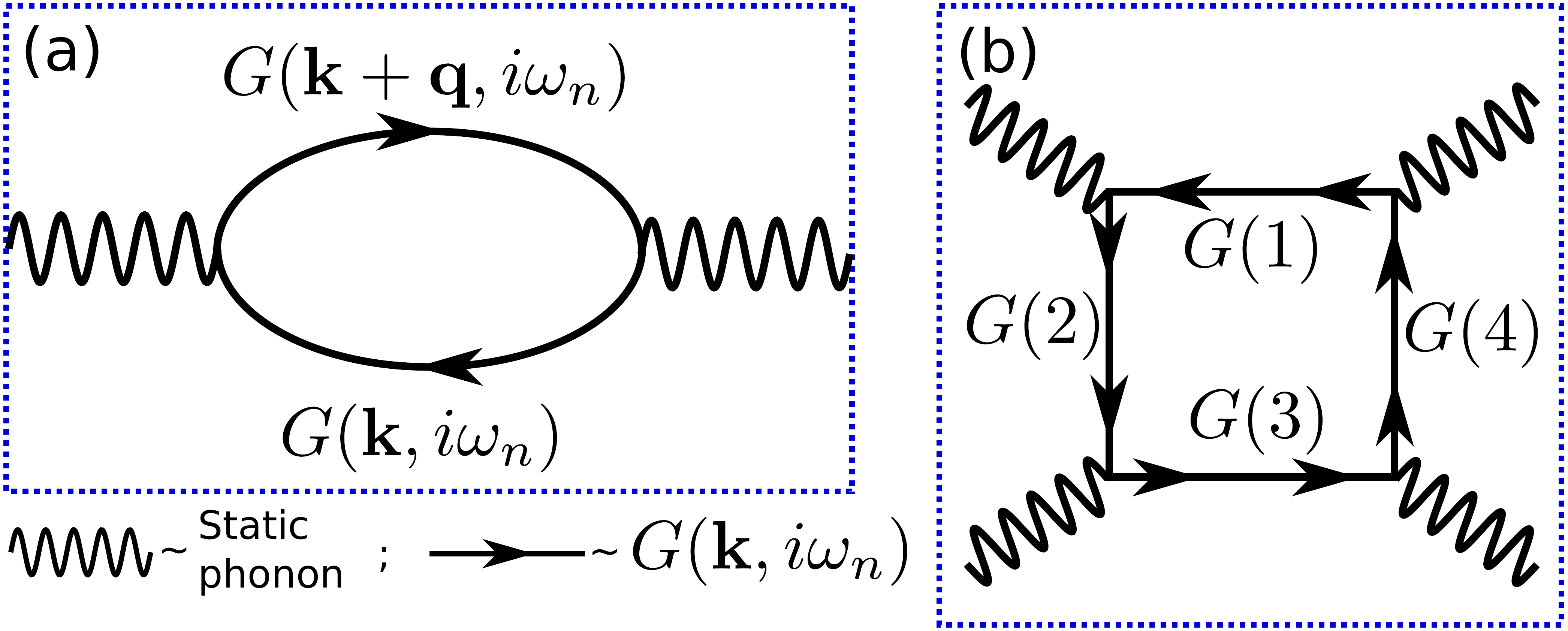}
\caption{(a) The second order Feynman diagram for the linked cluster term  $U_2$ as in Eq.~\ref{Coefficient:a}. The phonon frequency is set at zero. (b) The fourth order Feynman diagram for the linked cluster term $U_4$ as in Eq.~\ref{Coefficient:b} with static phonons. All possible combinations of the momenta, constrained by the momentum conservation at the vertices, are considered.}\label{fig:b1gmode}
\end{figure}
where $\mathcal{F}_0$ is the free energy of the electronic system resulting from the Hamiltonian $H_0$. The second term in Eq.~(\ref{Free_energy}) follows from the phonon Hamltonian $H_{\mathrm{Ph}}$. Finally, the linked cluster coefficients $U_l$ are obtained from the \textit{el-ph} Hamiltonian $H_{\mathrm{EP}}$ as (see Ref.~\cite{Mahan2011} for a detailed discussion about linked cluster expansion)
\begin{equation}\label{coefficient}
U_l = \frac{(-1)^l}{l \beta} \int_{0}^{\beta} d\tau_1 \int_{0}^{\beta} d\tau_2 .. \int^{\beta}_{0} \Braket{T_{\tau} H_{\mathrm{EP}}(\tau_1) H_{\mathrm{EP}}(\tau_2)...H_{\mathrm{EP}}(\tau_l)}_{connected}.
\end{equation}
The notation $\Braket{..}_{connected}$ in Eq.~(\ref{coefficient}) refers to all the distinct connected diagrams in the perturbative expansion of the mean-field Hamiltonian in Eq.~(\ref{Mean_field}). Restricting the choice of the ansatz wavevector $\vq$ to only fundamental wavevectors $\pm \vq^*, \pm \overline{\vq}^*$, we expand the last term in Eq.~(\ref{Free_energy}) up to fourth order to obtain the  coefficients 
\begin{subequations}\label{Coefficient}
\begin{align}
\label{Coefficient:a}
& U_2 = \frac{1}{2\beta}\sum_{\vk}\sum_{\vq = \pm  \vq^*,\pm \overline{\vq}^*} \sum_{i\omega_n} |\Delta(\vq) \tilde{\mathsf{g}}(\vq;\vk)|^2 G^0(\vk+\vq, i\omega_n) G^0(\vk, i\omega_n), \\
\label{Coefficient:b}
& \: \: \:  U_4 = \frac{1}{4\beta} \sum_{\vk_{i}} \sum_{\vq_i=\pm \vq^*,\pm \overline{\vq}^*} \sum_{i\omega_n} \Bigg[ \sideset{}{'} \prod_{i=1}^4 \Delta(\vq_i) \tilde{\mathsf{g}}(\vq_i;\vk_i) G^0(\vk_i+\vq_i, i\omega_n)\Bigg]
\end{align}
\end{subequations}
where we have rewritten the \textit{el-ph} Hamiltonian $H_{\mathrm{EP}}$ as
\begin{equation}\label{eq.1}
H_{\mathrm{EP}} = \sum_{\vk} \sum_{\vq =\pm \vq^*,\pm \overline{\vq}^*} \Delta(\vq)\tilde{\mathsf{g}}(\vq;\vk) c^{\dagger}_{\vk+\vq}c_{\vk}, 
\end{equation}
The quantity $\Delta(\vq)$ in Eq.~(\ref{eq.1})was defined in Eq.~(7) in the main text. The primed product in Eq.~(\ref{Coefficient:b}) implies momentum conservation at the vertices of Fig.~\ref{fig:b1gmode}(b), which in turn restricts the choices of the phonon scattering momenta $\vq_i$'s as 
\begin{align}\label{Constraint}
\vk_1 + \vq_1 = \vk_2,\; \vk_2 + \vq_2 = \vk_3, \; \vk_3 + \vq_3 = \vk_4, \vk_4 + \vq_4 = \vk_1.
\end{align}
Adding each term in Eq.~(\ref{Constraint}), we obtain the following constraint on the momenta $\vq_i$'s (in addition to the restriction $\vq_i\in (\pm \vq^*, \pm \overline{\vq}^*)$) as
\begin{equation}\label{phonon_constraint}
\vq_1 + \vq_2 + \vq_3 + \vq_4 =0.
\end{equation}
Similar constrains on the phonon scattering momenta $\vq_i$'s also arise for the odd order linked cluster terms. However, such conditions can only be satisfied for even order terms as  $\vq_i\in (\pm \vq^*, \pm \overline{\vq}^*)$ and consequently all the odd order cluster terms are zero. 

Performing the frequency summation in Eq.~(\ref{Coefficient}),  we obtain the fourth-order coefficient $\chi^{(4)}_{\vq}$ as (Eq.~(7) in the main text)
\begin{align}\label{chi_4_q}
\chi^{(4)}_{\vq} & = -\sum_{\vk}  \frac{1}{2}\frac{(\tilde{\mathsf{g}}(\vq;\vk))^4}{(\varepsilon_{\vk}-\varepsilon_{\vk'})^2} \left( \frac{2\{f(\varepsilon_{\vk'})-f(\varepsilon_{\vk})\}}{\varepsilon_{\vk}-\varepsilon_{\vk'}}+ \left( f'(\varepsilon_{\vk})+ f'(\varepsilon_{\vk'})\right)\right) -  \sum_{\vk} (\tilde{\mathsf{g}}(\vq;\vk))^2 (\tilde{\mathsf{g}}(\vq;\vk+\vq))^2 \cdot  \\ \nonumber
& \Bigg[ \frac{f(\varepsilon_{\vk})-f(\varepsilon_{\vk'})}{(\varepsilon_{\vk}-\varepsilon_{\vk''})(\varepsilon_{\vk}-\varepsilon_{\vk'})^2} - \frac{\beta f(\varepsilon_{\vk'})(f(\varepsilon_{\vk'})-1)}{(\varepsilon_{\vk}-\varepsilon_{\vk''})(\varepsilon_{\vk}-\varepsilon_{\vk'})} - \frac{f(\varepsilon_{\vk''})-f(\varepsilon_{\vk'})}{(\varepsilon_{\vk}-\varepsilon_{\vk''})(\varepsilon_{\vk''}-\varepsilon_{\vk'})^2} +  \frac{\beta f(\varepsilon_{\vk'})(f(\varepsilon_{\vk'})-1)}{(\varepsilon_{\vk}-\varepsilon_{\vk''})(\varepsilon_{\vk''}-\varepsilon_{\vk'})} \Bigg] .
\end{align}
The final fourth-order coefficient $\chi^{(4)}_{\vq^*;\overline{\vq}^*}$ is obtained as follows 
\begin{align}\label{chi_4_qq*}
\chi^{(4)}_{\vq^*;\overline{\vq}^*} & = -\sum_{\vk} (\tilde{\mathsf{g}}(\vq^*;\vk))^2 (\tilde{\mathsf{g}}(\overline{\vq}^*;\vk))^2 \Bigg[ \frac{f(\varepsilon_{\vk+\overline{\vq}^*})-f(\varepsilon_{\vk})}{(\varepsilon_{\vk+\overline{\vq}^*}-\varepsilon_{\vk+\vq^*})(\varepsilon_{\vk+\overline{\vq}^*}-\varepsilon_{\vk})^2} - \frac{\beta f(\varepsilon_{\vk})(f(\varepsilon_{\vk})-1)}{(\varepsilon_{\vk+\overline{\vq}^*}-\varepsilon_{\vk+\vq^*})(\varepsilon_{\vk+\overline{\vq}^*}-\varepsilon_{\vk})} \\ \nonumber
& - \frac{f(\varepsilon_{\vk+\vq^*})-f(\varepsilon_{\vk})}{(\varepsilon_{\vk+\overline{\vq}^*}-\varepsilon_{\vk+\vq^*})(\varepsilon_{\vk+\vq^*}-\varepsilon_{\vk})^2} + \frac{\beta f(\varepsilon_{\vk})(f(\varepsilon_{\vk})-1)}{(\varepsilon_{\vk+\overline{\vq}^*}-\varepsilon_{\vk+\vq^*})(\varepsilon_{\vk+\vq^*}-\varepsilon_{\vk})}\Bigg] - \sum_{\vk} (\tilde{\mathsf{g}}(\vq^*;\vk))^2 (\tilde{\mathsf{g}}(-\overline{\vq}^*;\vk))^2 \cdot \\ \nonumber
& \Bigg[ \frac{f(\varepsilon_{\vk-\overline{\vq}^*})-f(\varepsilon_{\vk})}{(\varepsilon_{\vk-\overline{\vq}^*}-\varepsilon_{\vk+\vq^*})(\varepsilon_{\vk-\overline{\vq}^*}-\varepsilon_{\vk})^2} - \frac{\beta f(\varepsilon_{\vk})(f(\varepsilon_{\vk})-1)}{(\varepsilon_{\vk-\overline{\vq}^*}-\varepsilon_{\vk+\vq^*})(\varepsilon_{\vk-\overline{\vq}^*}-\varepsilon_{\vk})} - \frac{f(\varepsilon_{\vk+\vq^*})-f(\varepsilon_{\vk})}{(\varepsilon_{\vk-\overline{\vq}^*}-\varepsilon_{\vk+\vq^*})(\varepsilon_{\vk+\vq^*}-\varepsilon_{\vk})^2} \\ \nonumber
& + \frac{\beta f(\varepsilon_{\vk})(f(\varepsilon_{\vk})-1)}{(\varepsilon_{\vk-\overline{\vq}^*}-\varepsilon_{\vk+\vq^*})(\varepsilon_{\vk+\vq^*}-\varepsilon_{\vk})}\Bigg] - \sum_{\vk} (\tilde{\mathsf{g}}(\vq^*;\vk))^2 (\tilde{\mathsf{g}}(\overline{\vq}^*;\vk+\vq^*))^2 \Bigg[ \frac{f(\varepsilon_{\vk})-f(\varepsilon_{\vk+\vq^*})}{(\varepsilon_{\vk}-\varepsilon_{\vk+\vq^*+ \overline{\vq}^*})(\varepsilon_{\vk}-\varepsilon_{\vk+\vq^*})^2} - \\ \nonumber
& \frac{\beta f(\varepsilon_{\vk+\vq^*})(f(\varepsilon_{\vk+\vq^*})-1)}{(\varepsilon_{\vk}-\varepsilon_{\vk+\vq^*+ \overline{\vq}^*})(\varepsilon_{\vk}-\varepsilon_{\vk+\vq^*})} - \frac{f(\varepsilon_{\vk+\vq^*})-f(\varepsilon_{\vk+\vq^*+ \overline{\vq}^*})}{(\varepsilon_{\vk}-\varepsilon_{\vk+\vq^*+ \overline{\vq}^*})(\varepsilon_{\vk+\vq^*}-\varepsilon_{\vk+\vq^*+ \overline{\vq}^*})^2}  +  \frac{\beta f(\varepsilon_{\vk+\vq^*})(f(\varepsilon_{\vk+\vq^*})-1)}{(\varepsilon_{\vk}-\varepsilon_{\vk+\vq^*+ \overline{\vq}^*})(\varepsilon_{\vk+\vq^*+ \overline{\vq}^*}-\varepsilon_{\vk+\vq^*})} \Bigg] \\ \nonumber
& - 2 \sum_{\vk} \frac{\tilde{\mathsf{g}}(\vq^*;\vk)\tilde{\mathsf{g}}(\vq^*;\vk+\overline{\vq}^*) \tilde{\mathsf{g}}(\overline{\vq}^*;\vk+\vq^*)\tilde{\mathsf{g}}(\overline{\vq}^*;\vk)}{(\varepsilon_{\vk+\vq^*}-\varepsilon_{\vk})(\varepsilon_{\vk +\vq^*+\overline{\vq}^*}-\varepsilon_{\vk+\overline{\vq}^*})} \Bigg[
\frac{f(\varepsilon_{\vk+\vq^*})-f(\varepsilon_{\vk +\vq^*+\overline{\vq}^*})}{\varepsilon_{\vk+\vq^*}-\varepsilon_{\vk +\vq^*+\overline{\vq}^*}} - \frac{f(\varepsilon_{\vk+\vq^*})-f(\varepsilon_{\vk+ \overline{\vq}^*})}{\varepsilon_{\vk+\vq^*}-\varepsilon_{\vk +\overline{\vq}^*}} \\  \nonumber
& -\frac{f(\varepsilon_{\vk})-f(\varepsilon_{\vk +\vq^*+\overline{\vq}^*})}{\varepsilon_{\vk}-\varepsilon_{\vk +\vq^*+\overline{\vq}^*}} + \frac{f(\varepsilon_{\vk})-f(\varepsilon_{\vk +\overline{\vq}^*})}{\varepsilon_{\vk}-\varepsilon_{\vk +\overline{\vq}^*}} \Bigg] - 2 \sum_{\vk} \frac{\tilde{\mathsf{g}}(\vq^*;\vk)\tilde{\mathsf{g}}(\vq^*;\vk-\overline{\vq}^*)\tilde{\mathsf{g}}(\overline{\vq}^*;\vk-\vq^*)\tilde{\mathsf{g}}(\overline{\vq}^*;\vk+\vq^*-\overline{\vq}^*)}{(\varepsilon_{\vk+\vq^*}-\varepsilon_{\vk})(\varepsilon_{\vk +\vq^*-\overline{\vq}^*}-\varepsilon_{\vk-\overline{\vq}^*})} \cdot  \\ \nonumber
&\Bigg[ \frac{f(\varepsilon_{\vk+\vq^*})-f(\varepsilon_{\vk +\vq^*-\overline{\vq}^*})}{\varepsilon_{\vk+\vq^*}-\varepsilon_{\vk +\vq^*-\overline{\vq}^*}}  - \frac{f(\varepsilon_{\vk+\vq^*})-f(\varepsilon_{\vk-\overline{\vq}^*})}{\varepsilon_{\vk+\vq^*}-\varepsilon_{\vk -\overline{\vq}^*}}-\frac{f(\varepsilon_{\vk})-f(\varepsilon_{\vk +\vq^*-\overline{\vq}^*})}{\varepsilon_{\vk}-\varepsilon_{\vk +\vq^*-\overline{\vq}^*}} + \frac{f(\varepsilon_{\vk})-f(\varepsilon_{\vk -\overline{\vq}^*})}{\varepsilon_{\vk}-\varepsilon_{\vk -\overline{\vq}^*}} \Bigg].
\end{align}
The fourth-order coefficient $\chi^{(4)}_{\vq^*;\overline{\vq}^*}$ decides between the possible uniaxial or biaxial character of the emergent charge order. However, both its magnitude and the sign depend sensitively on the parameters specific to the system. Hence, one cannot make any general statements about the character of the pattern of the resulting charge modulation. 

\subsection{Renormalized charge susceptibility: Pseudogap phase}
\begin{figure}[t]
\centering
\includegraphics[width=0.39\linewidth]{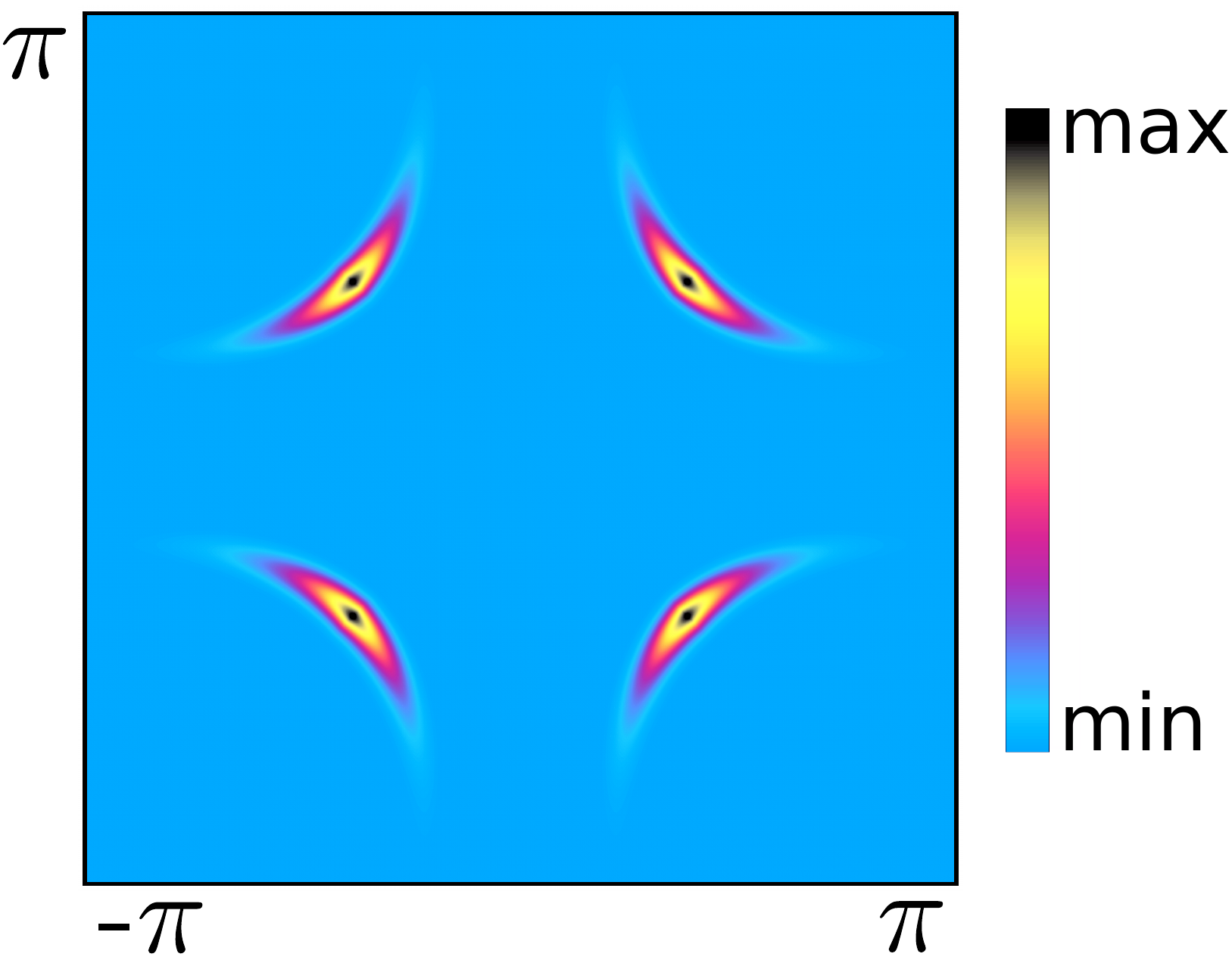}
\caption{The renormalized quasiparticle weight ansatz $Z_{\vk}$ along the Fermi surface at 10 $\%$ doping. The nodal points are maximally weighted whereas at the Brillouin zone faces the weight drops to zero.}\label{fig:fudge}
\end{figure}

 In this section, we derive the renormalized charge susceptibility as described in main text in Eq.~(11). We obtain the spectral function from Eq.~(9) (in the main text) as~\cite{PhysRevB.89.024507}

 \begin{equation}\label{spectral_function}
A(\vk, \omega)  = - \frac{1}{\pi} Im [G(\vk, \omega)] = Z_{\vk} \delta\left(\omega - \varepsilon_{\vk} \right) + \frac{1-Z_{\vk}}{2W} F(\omega),
\end{equation}
where the last term is our ansatz for $Im[G_{inc}]$. $W$ in Eq.~\ref{spectral_function} is the band-width of the anti-bonding band. The function $F(\omega)$ is defined as
\begin{equation}\label{ansatz}
F(\omega) = \left\{
\begin{array}{@{}rl@{}}
&1;\quad -W \leq \omega \leq W\\
\\
&0; \quad else   \\[2ex]
\end{array}
\right .
\end{equation}
Utilizing the spectral representation of the Green's function we obtain the renormalized static charge susceptibility as
\begin{align}\label{charge_susc}
\chi^R_{\vq} & = -2\sum_{\vk}|\mathsf{g}(\vq;\vk)|^2 \Bigg[ Z_{\vk} Z_{\vk'} \frac{f(\varepsilon_{\vk})-f(\varepsilon_{\vk'})}{i\omega_m + \varepsilon_\vk{-\varepsilon_{\vk'}}}  + \frac{Z_{\vk}(1-Z_{\vk'})}{2W} \int_{-W}^{W} d\omega' \frac{f(\varepsilon_{\vk})-f(\omega')}{i\omega_m + \varepsilon_{\vk} - \omega'} \\ \nonumber
& + \frac{Z_{\vk'}(1-Z_{\vk})}{2W} \int_{-W}^{W} d\omega \frac{f(\omega)-f(\varepsilon_{\vk'})}{i\omega_m + \omega - \varepsilon_{\vk'}} + \frac{(1-Z_{\vk'})(1-Z_{\vk})}{4W^2} \int_{-W}^{W} d\omega'd\omega \frac{f(\omega)-f(\omega')}{i\omega_m + \omega - \omega'} \Bigg]_{i\omega_m \mapsto \epsilon(=0) + i\delta }.
\end{align}
We now define the renormalized charge susceptibility $\chi^{(\mathsf{g}Z)}_{\vq}$ as following
\begin{equation}\label{renorm_susc}
\chi^{(\mathsf{g}Z)}_{\vq} = -2\gamma^2\sum_{\vk} Z_{\vk}Z_{\vk'}|\tilde{\mathsf{g}}(\vq;\vk)|^2 \Bigg[\frac{f(\varepsilon_{\vk})-f(\varepsilon_{\vk'})}{\varepsilon_{\vk}-\varepsilon_{\vk'}}\Bigg],
\end{equation}
where the last three terms in Eq.~\ref{charge_susc} have been neglected. We evaluate these terms at zero temperature and notice that all of them scale with the band-width $W$ as $(\log W)/W$. Hence, these terms are quantitatively negligible and the renormalized charge susceptibility is defined as in Eq.~(10) in the main text.

\begin{figure}[b]
\centering
\includegraphics[width=0.5\linewidth]{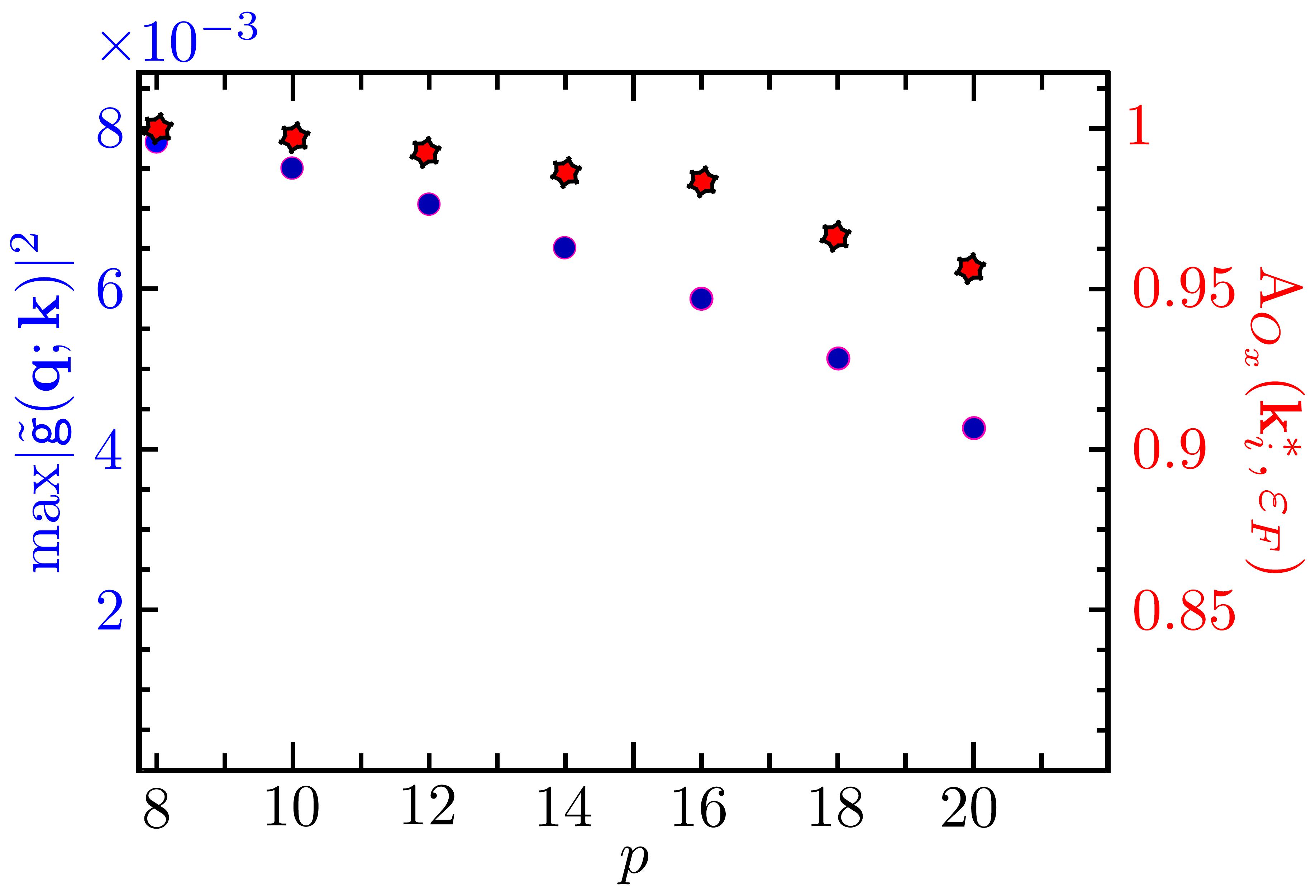}
\caption{The variation of the global maximum of the dimensionless \textit{el-ph} coupling $|\tilde{\mathsf{g}}(\vq;\vk)|^2$ (filled circles) and the maximum oxygen spectral weight on the Fermi surface at doping dependent initial momentum $\vk_i^*$ (filled diamonds) as a function of hole doping concentration $p$. Please note that we have normalized the maximum oxygen spectral weight at $8\%$ hole doping to 1, in comparison to the units for Fig.~2d in the main text.}\label{fig:doping}
\end{figure}
\subsection{Doping dependence of the \textit{el-ph} coupling}

In this section, we provide quantitative support for our reasoning on the expected dome-shaped region of the CDW in the phase diagram. First, we notice that the relevant initial electron momentum $\vk_i^*$, for which the \textit{el-ph} coupling $\mathsf{g}(\vq,\vk_i^*)$ attains its global maximum at $\vq^*$, varies by almost 5$\%$ with increasing the hole doping concentration (see Fig.~3 in the main text) \--- it moves closer to the nodal points on the Fermi surface. Additionally, the oxygen orbital content on the Fermi surfaces drops with hole doping as shown in Fig.~\ref{fig:doping}. The combined effects of the oxygen orbital content and phonon eigenvectors (see Eq.~5 in the main text) leads to an almost $50\%$ decrease of the squared \textit{el-ph} matrix element $|\mathsf{g}(\vq^*;\vk_i^*)|^2$between $p = 8\%$ and $p = 20 \%$.

\bibliographystyle{apsrev4-1}
\bibliography{Charge_cuprates}